\documentclass[amssymb,showkeys]{revtex4}

\usepackage{graphicx}% Include figure files
\usepackage{dcolumn}% Align table columns on decimal point
\usepackage{bm}% bold math
\usepackage[colorlinks]{hyperref}
\begin{document}

%\preprint{APS/123-QED}

\title{Asymptotic behavior of the node degrees   
in the ensemble average of adjacency matrix
}% Force line breaks with \\

\author{Yukio Hayashi}
\email{yhayashi@jaist.ac.jp}
\affiliation{Japan Advanced Institute of Science and Technology}
%\altaffiliation[Also at ]{Physics Department, XYZ University.}%Lines break automatically or can be forced with \\}
%\author{}
%\affiliation{Hitachi Co. Ltd.,\\

\date{\today}% It is always \today, today,

\begin{abstract}
Various important and useful quantities or measures that 
characterize the topological network structure are usually 
investigated for a network, then they are averaged 
over the samples.
In this paper, 
we propose an explicit representation by the beforehand averaged 
adjacency matrix over samples of growing networks 
as a new general framework for investigating the 
characteristic quantities. 
It is applied to some network models, 
and shows a good approximation of degree distribution 
asymptotically.
In particular, 
our approach will be applicable through the numerical calculations 
instead of intractable theoretical analysises, 
when the time-course of degree is a monotone increasing function 
like power-law or logarithm.
\end{abstract}

%\pacs{05.65.+b,89.65.-s,02.50.-r,05.10.-a,89.20.-a}
%\pacs05.90.+m, 02.50.-r, 02.60.-x, 05.65.+b}
% Systems obeying scaling law, Markov process, 
% Numerical simulation;solution of equations,
% Random walks and Levy flights, Self-organized systems
%05.65.+b Self-organized systems
%89.65.-s Social and economic systems
%02.50.-r Probability theory, stochastic orocesses, and statistics
%05.10.-a Computational methods in statistical physics and nonlinear dynamics
%89.20.-a Interdisciplinary applications of physics
%05.45.Df Fractals
%02.60.-x Numerical approximation and analysis


\maketitle

%\section{\label{sec:level1}First-level heading:\protect\\ The line
%break was forced \lowercase{via} \textbackslash\textbackslash}
%\subsection{\label{sec:level2}Second-level heading: Formatting}

\section{Introduction}
Many social, technological, and biological networks belong to a common 
scale-free (SF) structure \cite{Barabasi99a} 
which consists of many low degree nodes and 
a few high degree nodes called as hubs. 
The degree distribution follows a power-law, 
therefore an SF network has an extreme vulnerability against 
hub attacks \cite{Albert00}. 
In addition, these real networks are classified into 
assotative and disassortative networks \cite{Newman03a}. 
For examples, typical social networks, 
e.g. coauthorships and actor collaborations, 
are assotative, 
while typical technological and biological networks, 
e.g. Internet, World-Wide-Web, protein-interaction, and food webs, 
are disassortative.
In assotative networks, 
nodes with similar degrees tend to be connected, and thus 
positive degree-degree correlations appear. 
In disassotative networks, 
nodes with different degrees: low and high degrees 
tend to be connected, and thus 
negative degree-degree correlations appear. 

Through the above findings in network science, 
superior network theories 
and efficient algorithms have been developed 
for analyzing network topology and dynamics \cite{Newman10}.
However, studies for the cases with degree-degree correlations 
are not clear 
enough for successful analysises of topological structures 
and epidemics on networks except some percolation analysises. 
Recently, it has been numerically and theoretically found that 
an onion-like structure with positive degree-degree correlations 
gives the nearly optimal robustness against hub attacks in an SF 
network \cite{Herrmann11,Schneider11,Tanizawa12}. 

On the other hand, 
the average behavior of stochastically generated network
models or empirical data samples of real networks 
is discussed in many applications. 
Usually, 
some characteristic quantities such as degree distribution or 
clustering coefficient are investigated for a network, 
then their quantities are averaged over the samples of networks 
in which the existence of 
a generation rule (mechanism) of the networks is assumed. 
In this paper, 
we focus on the beforehand averaged network structure over 
samples, and calculate the degree distribution for several models 
of growing network with or without degree-degree correlations.
This representation will give a general framework for numerically 
investigating the characteristic topological quantities 
in growing networks.
Since our framework is supported by the interesting property of ordering 
that older nodes tend to have higher degrees in a randomly growing network
\cite{Callaway01}, a wide range of application to growing networks 
can be expected.

\section{Representation by the ensemble average of adjacency matrix}
We consider a set of growing networks in which 
a new node is added with probabilistic links to existing nodes 
in a network at each time step.
To study the average behavior of the stochastic processes in many 
samples, we use the ensemble average of adjacency matrix 
defined as follows.
\[
A(t) \stackrel{\rm def}{=} 
\left(\begin{array}{ccccccccc}
0      & 1      & a_{13}  & \ldots & a_{ij} & \ldots & a_{1i}  & \ldots & a_{1n} \\
1      & 0      & a_{23}  & \ldots & a_{2j} & \ldots & a_{2i}  & \ldots & a_{2n} \\
a_{31}  & a_{32}  & 0      & \ldots & a_{3j} & \ldots & a_{3i} & \ldots & a_{3n} \\
\vdots & \vdots & \vdots & \ddots & \vdots & \vdots & \vdots & \vdots & \vdots \\
a_{j1}  & a_{j2}  & a_{3j}  & \ldots & 0      & \ldots & a_{ji} & \ldots & a_{jn} \\
\vdots & \vdots & \vdots & \vdots & \vdots & \ddots & \vdots & \vdots & \vdots \\
a_{i1}  & a_{i2}  & \ldots & \ldots  & a_{ij} & \ldots & 0      & \ldots & a_{in} \\
\vdots & \vdots & \vdots & \vdots & \vdots & \vdots & \vdots & \ddots & \vdots \\    
a_{n1} & a_{n2}   & \ldots & \ldots & \ldots & \ldots & a_{ni}  & \ldots & 0 \\ 
\end{array}\right)
\]
Here, without loss of generality, 
we set connected two nodes as the minimum initial configuration: 
$a_{11} = a_{22} = 0$, $a_{12} = a_{21} = 1$, and the degree 
$k_{1}(0) = k_{2}(0) = 1$.
Note that 
at each time step the matrix is expanding with the inverse $L$ shape 
of elements 
$a_{1n}, a_{2n}, \ldots, a_{n-1n}$, $0$, $a_{nn-1}, \ldots, a_{n2}, a_{n1}$ 
at the right-bottom corner.
The diagonal element $a_{ii}$ is always $0$ due to no self-loop at 
each node $i$.
Other elements are $0 \leq a_{ij} \leq 1$, $i \neq j$, 
as the average number of links from $i$ to $j$ 
over the samples of networks.
The value of each element is defined in order according to the 
passage of time.
We assume that there are no adding links between existing nodes 
$i, j < n$ at any time $t=n-2$.
Only links between selected old nodes and new node $n$ 
are added in a growing network.

In the sample-based description, 
the ensemble average of adjacency matrix is 
\[
  A(t) = \frac{A^{(1)}(t) + A^{(2)}(t) + \ldots + 
    A^{(k)}(t) + \ldots + A^{({\cal N}_{s})}(t)}{{\cal N}_{s}},
\]
where $A^{(k)}(t)$ denotes an adjacency matrix of the $k$-th 
sample whose elements are $a_{ij}^{(k)} = 1$ or $0$ 
corresponded to the connection or disconnection 
from $i$ to $j$, but $a_{ij}^{(k)} = 0$ or undefined 
for $i, j > t+2$ because of the size 
$n=t+2$ at time $t$. 
${\cal N}_{s}$ denotes the number of samples.
We remark that 
an adjacency matrix is fundamental and important 
because it includes the necessary and sufficient information 
about a network structure and is useful for a mathematical treatment.

In this explicit representation of general framework, 
for each $i$-th node, $i = 1, 2, \ldots, n-1$, 
the (out-)degree is updated from time $t-1$ to $t$, 
\begin{equation}
  k_{i}(t) = k_{i}(t-1) + a_{in}.
\label{eq_DE_general_ki}
\end{equation}
The (out-)degree of $n$-th node added at time $t=n-2$ 
is defined by the sum of links from node $n$ to nodes $i$, 
\begin{equation}
  k_{n}(t) \stackrel{\rm def}{=} \sum_{i=1}^{n-1} a_{ni}.
\label{eq_DE_general_kn}
\end{equation}

We should remark that 
the iterative calculations of 
Eqs. (\ref{eq_DE_general_ki})(\ref{eq_DE_general_kn}) 
are equivalent to the averaged values over the samples 
after calculating the degrees for each sample of 
the networks at time $t$. 
With this equivalence in mind, 
we investigate the asymptotic behavior of $k_{i}(t)$
for a large $t$.
We note that $k_{i}(t)$ is a monotone non-decreasing function 
of time $t$ because of 
$k_{i}(t) - k_{i}(t-1) = a_{in} \geq 0$ 
from Eq. (\ref{eq_DE_general_ki}) in growing networks.

\section{Asymptotic behavior of the node degrees}
As examples, 
we apply the ensemble average of adjacency matrix 
to some network models.
However, our approach is applicable to other 
growing networks especially in a wide class, e.g. 
with approximately power-law or exponential degree distribution.
In the following, 
we assume that each link is undirected: $a_{ij} = a_{ji}$.

\subsection{Babar\'{a}si and Albert model}
Since the continuous-time approximation of Eq.(\ref{eq_DE_general_ki}) 
is generally 
\[
  \frac{d k_{i}(t)}{d t} = a_{in}, 
\]
our approach is applicable to the Babar\'{a}si and Albert (BA) model
\cite{Barabasi99b} as follow.

preferential attachment: 
\[
  a_{ni} = \frac{m \times k_{i}(t-1)}{\sum_{l} k_{l}(t-1)} 
  \approx \frac{k_{i}(t-1)}{2 t},
\]

uniform attachment: 
\[
  \forall i \;\;\; a_{ni} = \frac{m}{m_{0} + t-1},
\]
where $m_{0}$ denotes the initial number of nodes, 
and $m$ denotes the number of adding links at each time step.

For the two cases of preferential and uniform attachments, 
$k_{i}(t) = m \sqrt{t/t_{i}}$ and 
$k_{i}(t) = m (\log(m_{0}+t-1) - \log(m_{0}+t_{i}-1) + 1)$ 
have been derived with the corresponding degree distributions 
$p(k) \sim k^{-3}$ and $p(k) \sim e^{-k}$, respectively 
\cite{Barabasi99b}. 
The analysis in BA model 
is based on the invariant ordering property of 
degrees in which older nodes get more links averagely at any time 
in the growth of network. 
Under the invariant ordering property, 
our approach can be regarded as an extension of mathematical treatment 
in the BA model 
%including another version of the difference equation (\ref{eq_DE_general_ki})
through the representation by the ensemble average of 
adjacency matrix over samples of growing networks.

\subsection{Duplication-divergence model}
\begin{figure}[htp]
 \begin{minipage}[htb]{0.47\textwidth} 
   \centering
   \includegraphics[height=50mm]{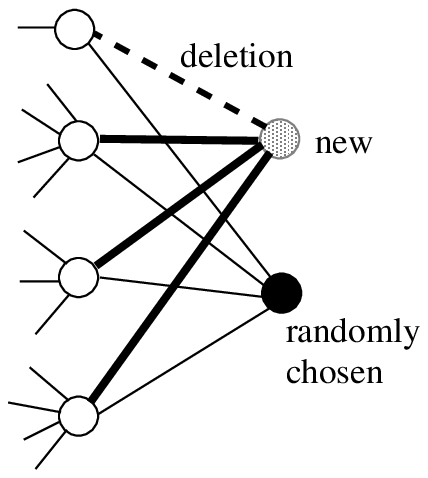}
     \begin{center} (a) D-D model \end{center}
 \end{minipage} 
 \hfill 
 \begin{minipage}[htb]{0.47\textwidth} 
   \centering
   \includegraphics[height=50mm]{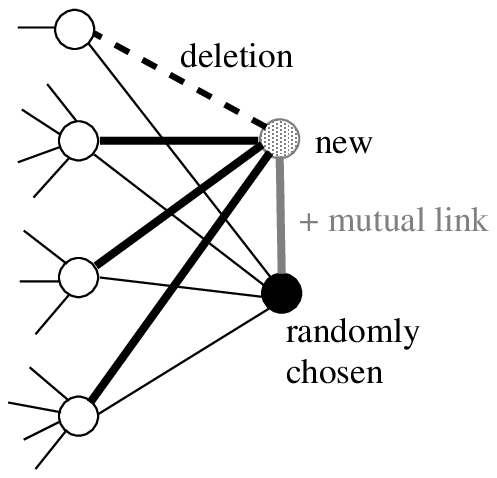}
     \begin{center} (b) Copying model \end{center}
 \end{minipage} 
\caption{Basic linking processes at each time step 
from a new node to 
a randomly chosen node and/or the neighbor nodes 
in the growing networks. 
The thick bold lines show adding links, and 
the dashed line shows a deleted one.
The thin lines show already existing links at that time.}
\label{fig_basic_process}
\end{figure}

We preliminary introduce 
a duplication-divergence (D-D) model \cite{Satorras03,Sole02} 
without mutations of random links between existing nodes, 
whose generation 
mechanism is known as fundamental in biological protein-protein 
interaction networks. 
In the D-D model, at each time step, 
a new node is added and links to neighbor nodes of a uniformly 
randomly chosen node (see Figure \ref{fig_basic_process}(a)). 
Some duplication links are deleted with probability $\delta > 0$.
Here, no mutations are to simplify the discussion and to connect 
to the next subsection.
Although the degree distribution in the D-D model 
can be approximately analyzed by the approach of 
mean-field-rate equation \cite{Satorras03,Sole02}, 
we show the applicability of our approach 
to the D-D model in order to extend it to more general networks.
Moreover, in the next section, 
we reveal that older nodes tend to have higher degrees 
\cite{Callaway01} in the D-D model, 
whose ordering of degree $k_{i}(t)$ for node index $i$ 
was not found from the above approach.

Since the $n$-th new node links to the neighbor node $i$ 
of a chosen node $j \neq i$ from existing nodes 
$1 \leq i, j \leq n-1$ in a network of D-D model, we have 
\begin{equation}
  a_{ni} = \frac{(1-\delta) \sum_{j=1}^{n-1} a_{ji}}{n-1} 
  = \frac{1}{n-1} (1 - \delta) k_{i}(t-1),
\label{eq_a_ni_D-D}
\end{equation}
where we use the uniform selection probability $1/(n-1)$ 
of each node $j$ and the no-deletion rate $1-\delta$ 
for linking to the neighbor nodes.

From Eqs. (\ref{eq_DE_general_ki})(\ref{eq_DE_general_kn})
(\ref{eq_a_ni_D-D}), we obtain 
\begin{equation}
  k_{i}(t) = \left( 1 + \frac{1-\delta}{n-1} \right) k_{i}(t-1), 
\label{eq_DE_D-D_ki}
\end{equation}
\begin{equation}
  k_{n}(t) = \frac{1-\delta}{n-1} \sum_{i=1}^{n-1} k_{i}(t-1).
\label{eq_DE_D-D_kn}
\end{equation}
By applying Eq. (\ref{eq_DE_D-D_ki}) recursively, we derive 
\begin{eqnarray}
  k_{i}(t) & = & \left( \frac{n-1 + (1-\delta)}{n-1} \right) 
  \left( \frac{n-2 + (1-\delta)}{n-2} \right) \ldots 
  \left( \frac{i + (1-\delta)}{i} \right) k_{i} \nonumber \\
  & = & \frac{\Gamma(n + (1-\delta))/\Gamma(i+(1-\delta))}{
    \Gamma(n)/\Gamma(i)} k_{i} \sim (t+1)^{1-\delta}. \nonumber
\end{eqnarray}

\begin{figure}[htp]
 \begin{minipage}{0.47\textwidth} 
   \centering
   \includegraphics[height=71mm,angle=-90]{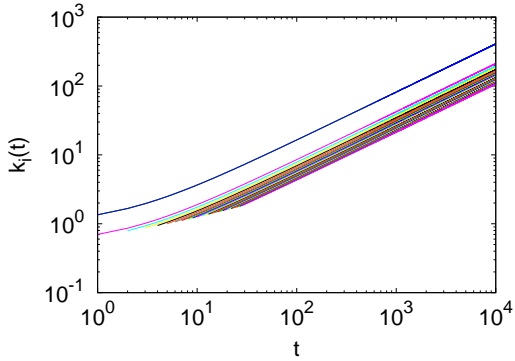}
     \begin{center} (a) $\delta=0.3$ \end{center}
 \end{minipage} 
 \hfill 
 \begin{minipage}{0.47\textwidth} 
   \centering
   \includegraphics[height=71mm,angle=-90]{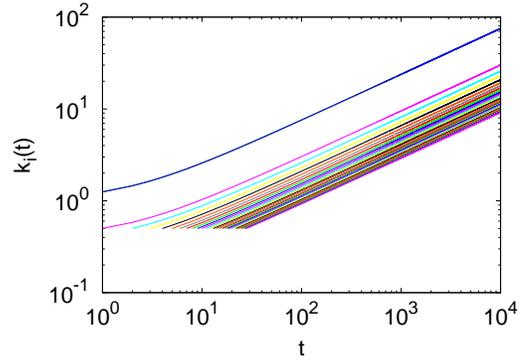}
     \begin{center} (b) $\delta=0.5$ \end{center}
 \end{minipage} 
 \hfill  
 \begin{minipage}{0.47\textwidth} 
   \centering
   \includegraphics[height=71mm,angle=-90]{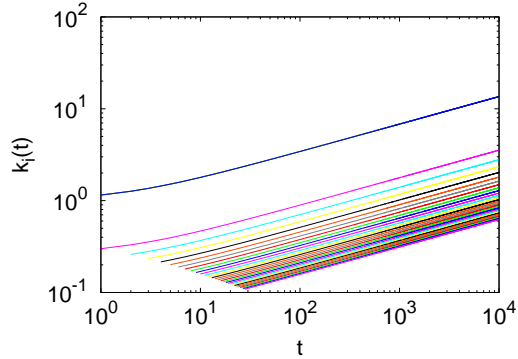}
     \begin{center} (c) $\delta=0.7$ \end{center}
 \end{minipage} 
\caption{Parallel curves of time-course $k_{i}(t)$ of degree
in the D-D model with a rate $\delta$ of link deletion. 
The color lines from top to bottom correspond to the nodes
$i=1$ or $2, 3, 4, \ldots, 30$. Note that each node $i$ is born at time 
$t_{i}=i-2$.}
\label{fig_para_D-D}
\end{figure}

In addition, the continuous-time approximation of Eq. (\ref{eq_DE_D-D_ki}) 
is 
\[
  \frac{d k_{i}(t)}{dt} = a_{ni} = \frac{1-\delta}{t+1} k_{i}(t).
\]
From the separation of variables method, 
we obtain the solution 
\[
  k_{i}(t) = (t+1)^{1-\delta} \sim t^{1-\delta}.
\]
When we denote the initial degree $k_{i}$ 
at the inserted time $t_{i}$ for a node $i$, 
the above solution is rewritten as 
\[
  k_{i}(t) \approx k_{i} \left( \frac{t}{t_{i}} \right)^{1-\delta}.
\]
From the existence of parallel curves shown in Fig. \ref{fig_para_D-D}, 
the ordering of degrees 
\begin{equation}
  k_{n}(t) < k_{n-1}(t) < \ldots < 
  k_{3}(t) < k_{2}(t)=k_{1}(t), \label{eq_ordering}
\end{equation}
is not changed. In other word, 
older nodes get more links averagely.
Thus, we obtain 
\[
  p(k_{i}(t) < k) = 
  p\left(t_{i} > \left( \frac{k_{i}}{k} \right)^{1/(1-\delta)} t \right)
  = \left( 1 - \frac{k_{i}^{1/(1-\delta)}}{k^{1/(1-\delta)}} \right) 
  \frac{t}{N_{0} + t}, 
\]
\begin{equation}
  p(k) = \frac{\partial p(k_{i}(t) < k)}{\partial k} 
  \sim k^{-(1 + \frac{1}{1-\delta})},
\label{eq_approx_pk}
\end{equation}
where $N_{0} + t$ is the number of nodes at time $t$, 
and $N_{0}$ denotes the initial number of nodes. 
In the tail of degree distribution, 
the exponent of power-law is $1 + \frac{1}{1-\delta}$ asymptotically. 
Note that the slightly different exponent $1 + \frac{1}{1-\delta}$ to 
the conventional approximation \cite{Kim02,Satorras03} is not strange, 
since the mutations are necessary for their D-D models.

\begin{figure}[htp]
 \begin{minipage}{0.47\textwidth} 
   \centering
   \includegraphics[height=67mm,angle=-90]{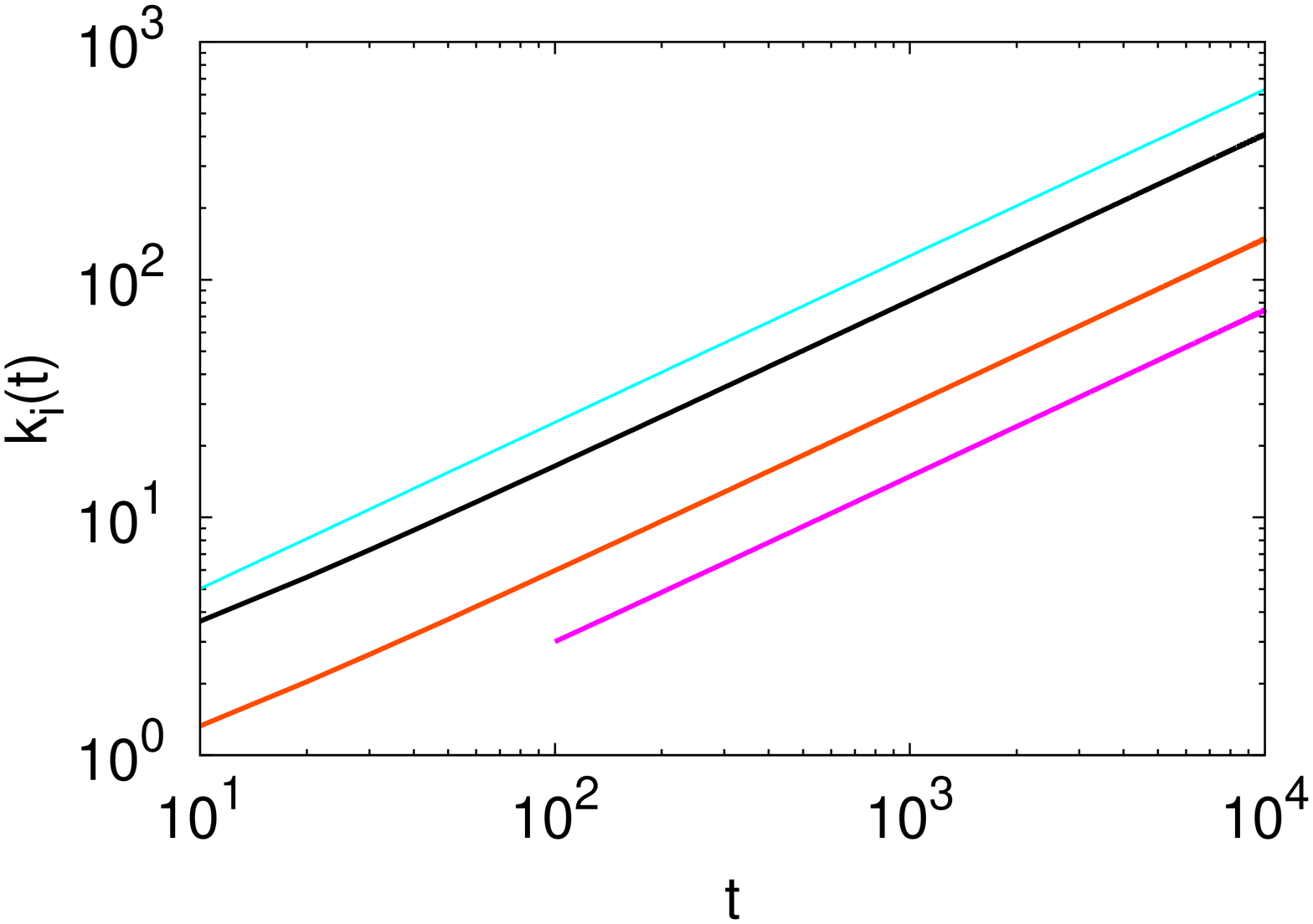}
     \begin{center} (a) $\delta=0.3$ \end{center}
 \end{minipage} 
 \hfill 
 \begin{minipage}{0.47\textwidth} 
   \centering
   \includegraphics[height=67mm,angle=-90]{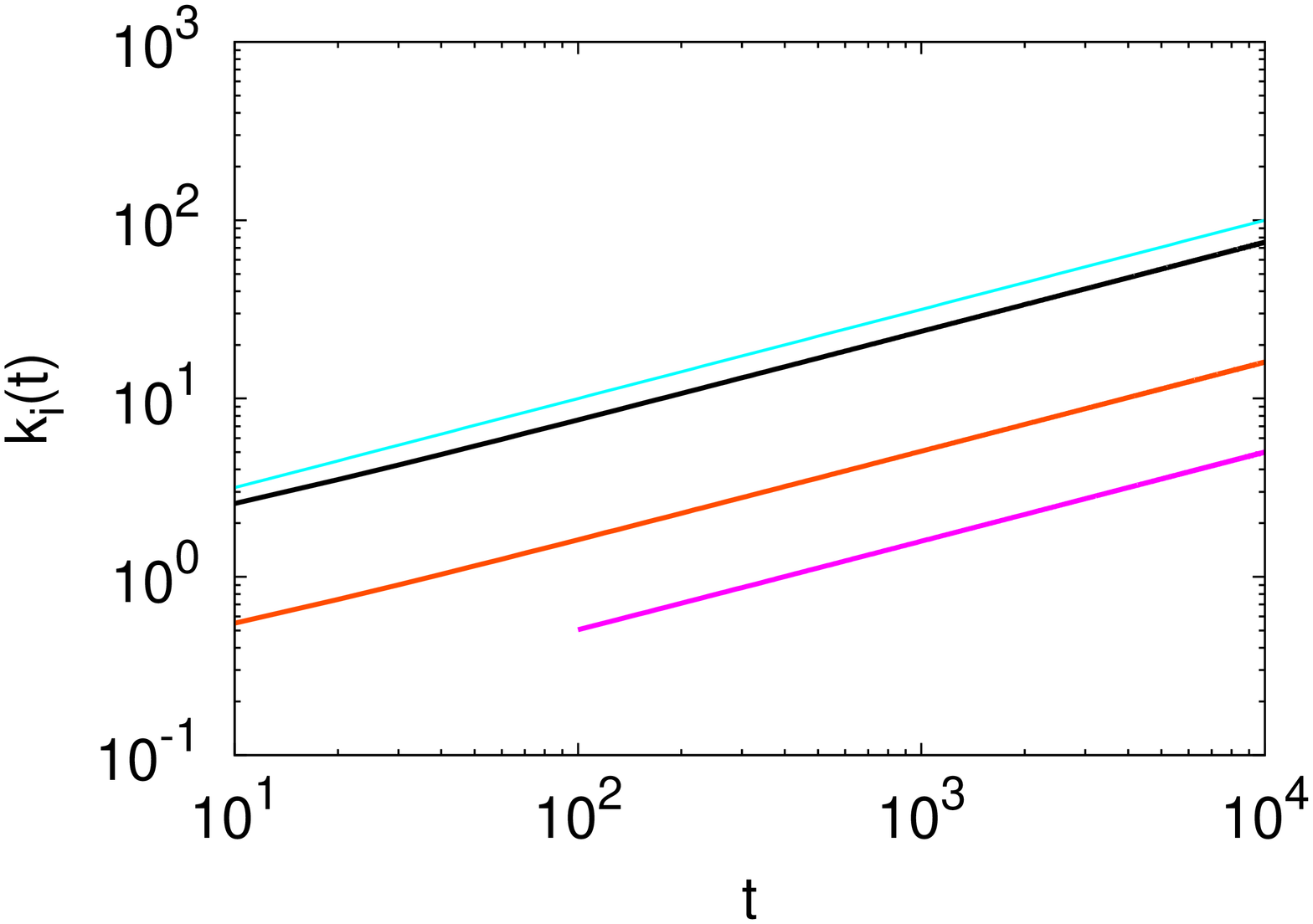}
     \begin{center} (b) $\delta=0.5$ \end{center}
 \end{minipage} 
 \hfill 
 \begin{minipage}{0.47\textwidth} 
   \centering
   \includegraphics[height=67mm,angle=-90]{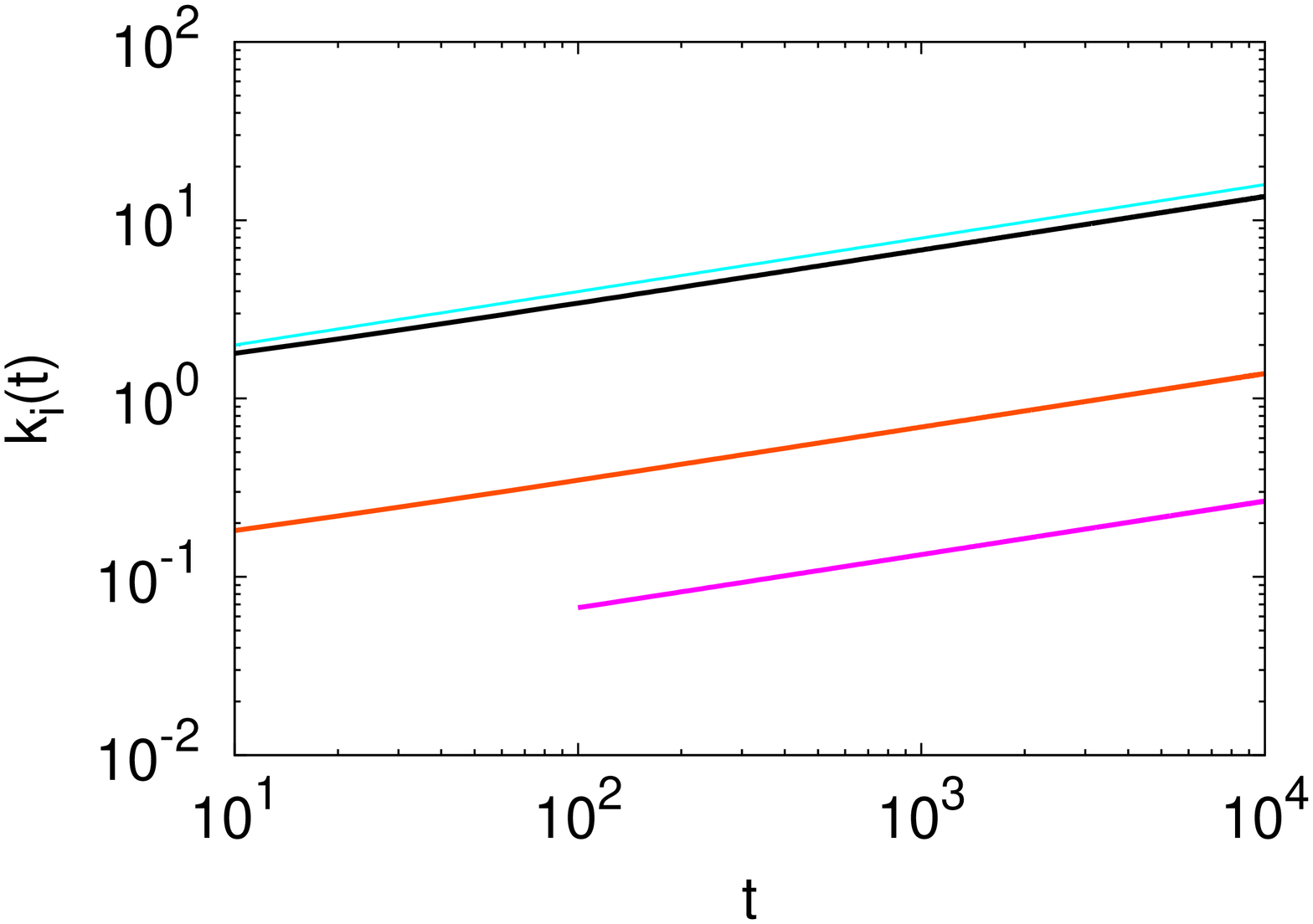}
     \begin{center} (c) $\delta=0.7$ \end{center}
 \end{minipage} 
 \hfill 
 \begin{minipage}{0.47\textwidth} 
   \centering
   \includegraphics[height=67mm,angle=-90]{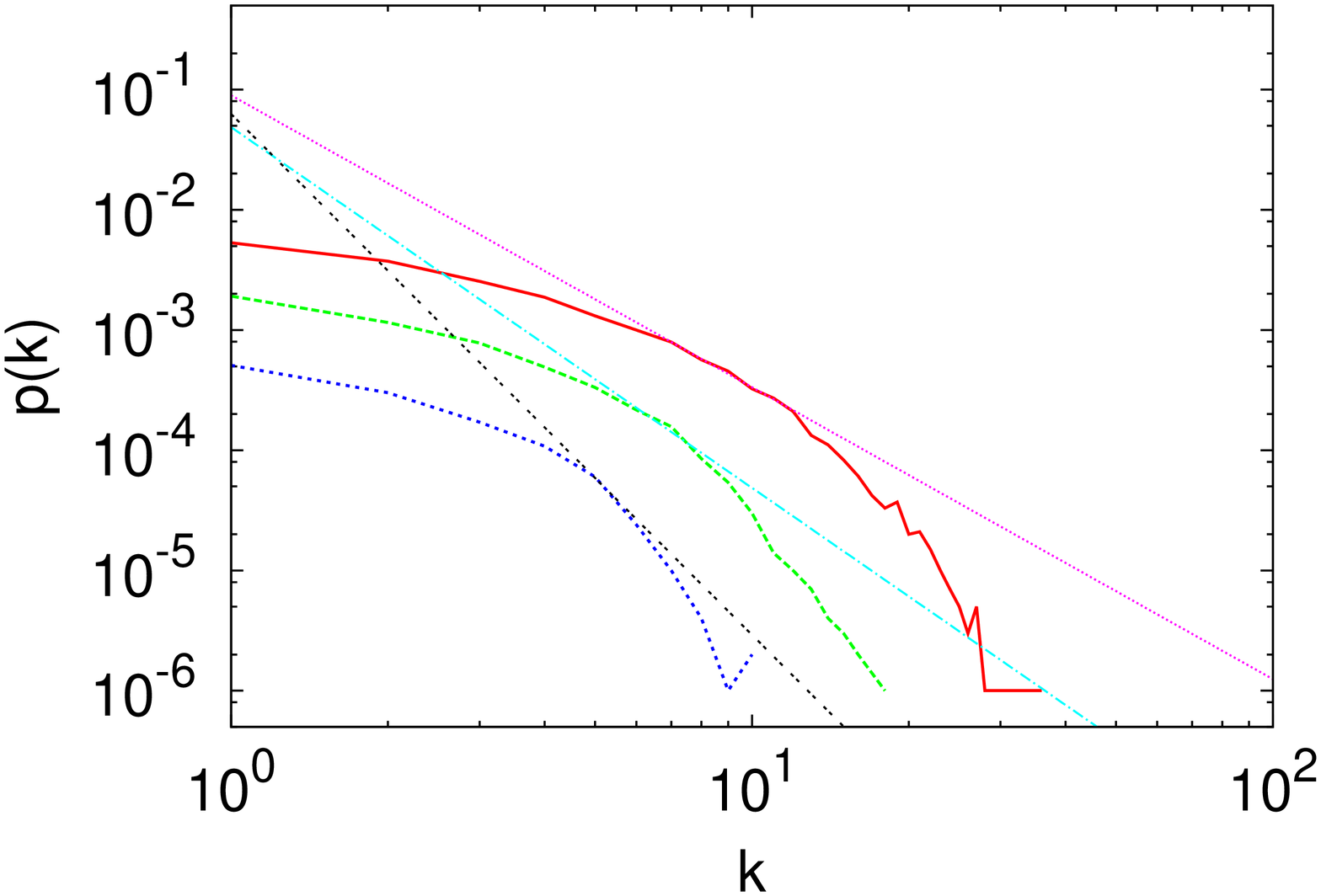}
     \begin{center} (d) $p(k)$  \end{center}
 \end{minipage} 
\caption{Results for the D-D model without mutations. 
(a)-(c) Time-courses $k_{i}(t)$ of degree for node $i=1, 10,$ and $100$ 
shown by black, orange, and magenta lines, respectively, 
at the typical values of rate $\delta$ of link deletion.
The cyan line guides the estimated slope $1 - \delta$ of 
$k_{i}(t) \sim t^{1 - \delta}$. 
(d) Degree distribution $p(k)$ in the cases of 
$\delta = 0.3, 0.5$, and $0.7$ shown by red, green, and blue lines, 
respectively.
The dotted magenta, cyan, and gray lines show the estimated power-law
distributions $k^{-\left(1+\frac{1}{1-\delta}\right)}$.}
\label{fig_D-D}
\end{figure}

Figure \ref{fig_D-D}(a)-(c) shows the time-course of 
$k_{i}(t) \sim t^{1-\delta}$ in the case of 
$\delta = 0.3$, $0.5$, and $0.7$, respectively, 
averaged over $100$ samples. 
The black, orange, and magenta lines are the numerical results 
of Eqs. (\ref{eq_DE_D-D_ki})(\ref{eq_DE_D-D_kn})
for the node $i = 1$, $10$, and $100$.
The cyan line guides the estimated slope of 
$1-\delta$ in log-log plot. 
In Fig. \ref{fig_D-D}(d), 
the red, green, and blue lines show the degree distributions for 
$\delta = 0.3$, $0.5$, and $0.7$, respectively, at the size 
$n =10^{4}$.
The magenta, cyan, and gray dashed lines guides the corresponding 
slopes of $1 + \frac{1}{1-\delta}$ for these $\delta$.

\subsection{Copying model}
A modification \cite{Hayashi14}
of the D-D model \cite{Satorras03,Sole02} 
by adding a mutual link between a new node and a randomly chosen node
has been proposed. 
The mutual link contributes to avoid the singularity called as 
non-self-averaging even for $\delta < 1/2$ without 
mutations \cite{Hayashi14}.
The growing network is constructed 
as shown in Fig. \ref{fig_basic_process}(b). 
It is referred to as copying model.
At each time step, a new node is added. 
The new node links to a uniformly randomly chosen node, 
and to its neighbor nodes with probability $1-\delta$.

In the copying model, we have 
\begin{equation}
  a_{ni} = \frac{1 + (1-\delta) \sum_{j=1}^{n-1} a_{ji} }{n -1} 
  = \frac{1 + (1-\delta) k_{i}(t-1)}{n-1}, 
\label{eq_a_ni_copying}
\end{equation}
since the $n$-th new node links to a uniformly randomly 
chosen node $i$ 
and to the neighbor node $i$ when other node $j$ is chosen
from existing nodes 
$1 \leq i, j \leq n-1$ in the network.
These effects are in the first term $1/(n-1)$ and 
the second term $(1-\delta) \sum_{j=1}^{n-1} a_{ji} / (n-1)$ 
in Eq. (\ref{eq_a_ni_copying}).

From Eqs. (\ref{eq_DE_general_ki})(\ref{eq_DE_general_kn})
(\ref{eq_a_ni_copying}), 
we obtain 
\begin{equation}
  k_{i}(t) 
  = \frac{1}{n-1} + \left( 1 + \frac{1-\delta}{n-1} \right) k_{i}(t-1), 
\label{eq_DE_mut_ki}
\end{equation}
\begin{equation}
  k_{n}(t) 
  = 1 + \frac{1-\delta}{n-1} \sum_{i=1}^{n-1} k_{i}(t-1).
\label{eq_DE_mut_kn}
\end{equation}

By applying Eq. (\ref{eq_DE_mut_ki}) recursively, 
we derive 
\begin{eqnarray}
  k_{i}(t) & = & \frac{1}{t+1} 
  + \left(\frac{t + 1 + (1-\delta)}{t+1} \right)
  \left( 
  \frac{1}{t} + \left( \frac{t+(1-\delta)}{t} k_{i}(t-2) \right) 
  \right) \nonumber \\
  & = & \frac{1}{t+1} + \frac{t + 1 + (1-\delta)}{(t+1)t} + \ldots + 
  \frac{\Gamma(t+2+(1-\delta))/\Gamma(i)}{\Gamma(t+2)/\Gamma(i)} k_{i}(t_{
i}) \nonumber \\
  & \approx & O(1/t) + (t+1)^{1-\delta}, 
\label{eq_DE_sol}
\end{eqnarray}
where we use $n-1 = t+1$ 
and the Staring formula $\Gamma(x+1) \approx x^{x} e^{-x}$.

In particular, by the mathematical induction, 
we confirm that the case of $\delta = 0$ generates a sequence 
of the complete graphs with 
$k_{i}(t) = t+1 = n-1$ links at every node of 
$i = 1, 2, \ldots, n$.
First,   
$k_{1}(1) = k_{2}(1) = 2$ and $k_{3}(1) = 2$
are obvious. 
Next, we assume 
$k_{i}(t-1) = (t-1) + 1 = n-2$, 
from Eqs.(\ref{eq_DE_mut_ki})(\ref{eq_DE_mut_kn}) 
we derive 
\begin{eqnarray*}
  k_{i}(t) & = & \frac{1}{n-1} 
  + \left( 1 + \frac{1}{n-1} \right)(n-2) 
  = \frac{1 + (n-1)(n-2) + (n-2)}{n-1} \\
  & = & \frac{(n-1)^{2}}{n-1} = n-1, 
\end{eqnarray*}
\[
  k_{n}(t) = 1 + \frac{(n-1)(n-2)}{n-1} = n-1.
\]

\begin{figure}[htp]
 \begin{minipage}{0.47\textwidth} 
   \centering
   \includegraphics[height=71mm,angle=-90]{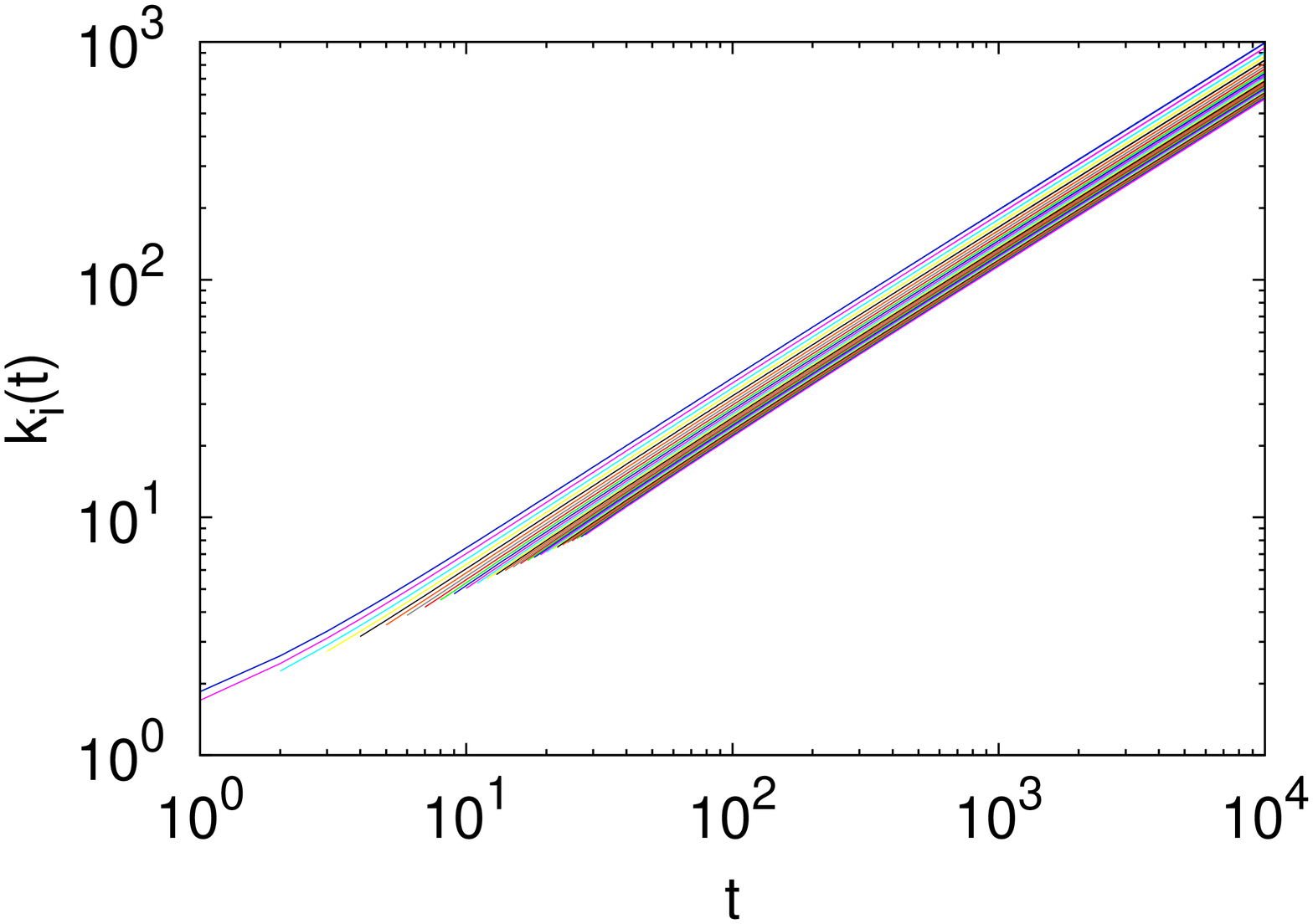}
     \begin{center} (a) $\delta=0.3$ \end{center}
 \end{minipage} 
 \hfill 
 \begin{minipage}{0.47\textwidth} 
   \centering
   \includegraphics[height=71mm,angle=-90]{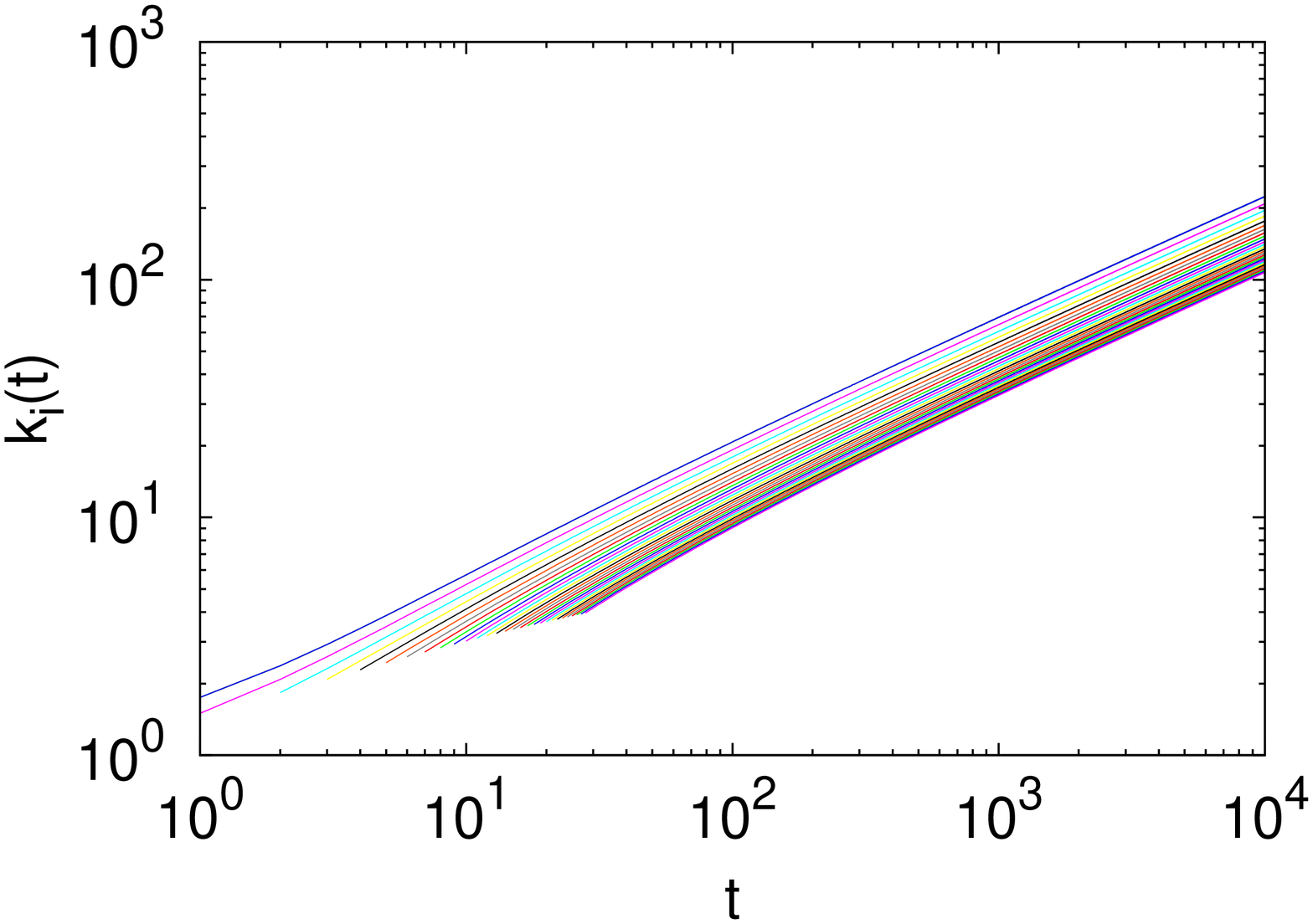}
     \begin{center} (b) $\delta=0.5$ \end{center}
 \end{minipage} 
 \hfill  
 \begin{minipage}{0.47\textwidth} 
   \centering
   \includegraphics[height=71mm,angle=-90]{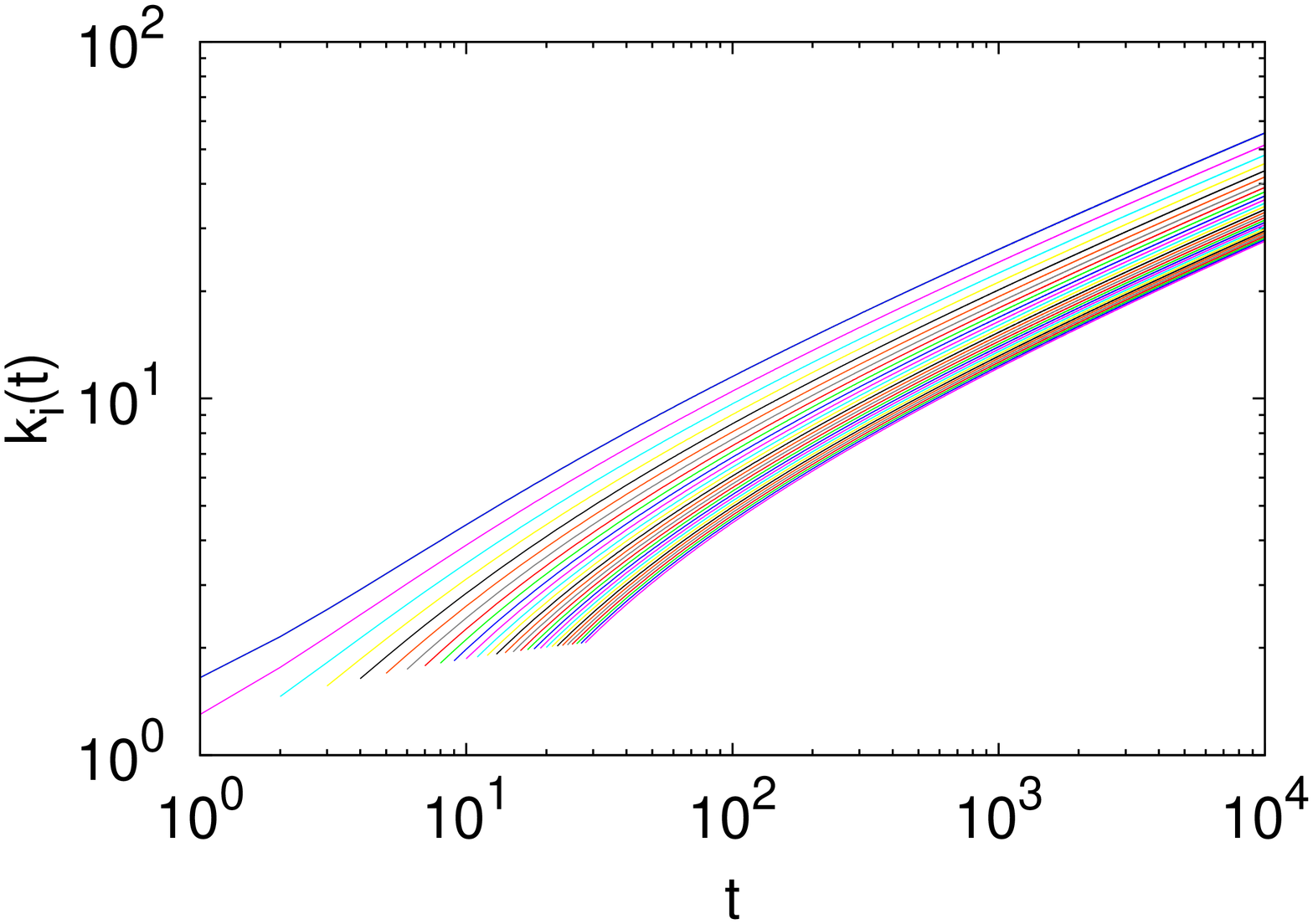}
     \begin{center} (c) $\delta=0.7$ \end{center}
 \end{minipage} 
\caption{Parallel curves of time-course $k_{i}(t)$ of degree 
in the copying model with a rate $\delta$ of link deletion. 
The color lines from top to bottom correspond to the nodes 
$i=1$ or $2, 3, 4, \ldots, 30$. Note that each node $i$ is born at time 
$t_{i}=i-2$.}
\label{fig_para_copying}
\end{figure}

On the other hand, 
the continuous-time approximation of Eq.(\ref{eq_DE_mut_ki}) is 
\[
  \frac{d k_{i}(t)}{dt} = a_{ni} = \frac{1}{t+1} 
  + \left( \frac{1-\delta}{t+1} \right) k_{i}(t).
\] 
Since this form is a 1st order linear differential equation 
$dk_{i} /dt + f(t) k_{i} = g(t)$, 
by applying the solution 
$e^{-\int f(t) dt} \left( e^{\int f(t) dt} g(t) dt + A \right)$, 
we obtain 
\[
  k_{i}(t) = (t+1)^{1-\delta} 
  \left( \int_{0}^{t} \frac{1}{(t+1)^{1+(1-\delta)}} dt + A \right)
  = A(t+1)^{1-\delta} - C' \sim t^{1-\delta},
\]
where $A$ and $C'$ are constants of integration, 
and we use 
$f(t) \stackrel{\rm def}{=} -(1-\delta)/(t+1)$ and 
$g(t) \stackrel{\rm def}{=} 1/(t+1)$. 
Note that the solution $t^{1-\delta}$ 
is only different by $O(1/t)$ to Eq.(\ref{eq_DE_sol}), 
and can be ignored for a large $t$.
As similar to the D-D model in subsection 3.2, 
from Eq. (\ref{eq_approx_pk}) 
under the invariant ordering (\ref{eq_ordering}) in the parallel 
curves shown as Fig. \ref{fig_para_copying}, 
the degree distribution asymptotically follows a power-law with 
the exponent $1 + \frac{1}{1-\delta}$.

\begin{figure}[htp]
 \begin{minipage}{0.47\textwidth} 
   \centering
   \includegraphics[height=71mm,angle=-90]{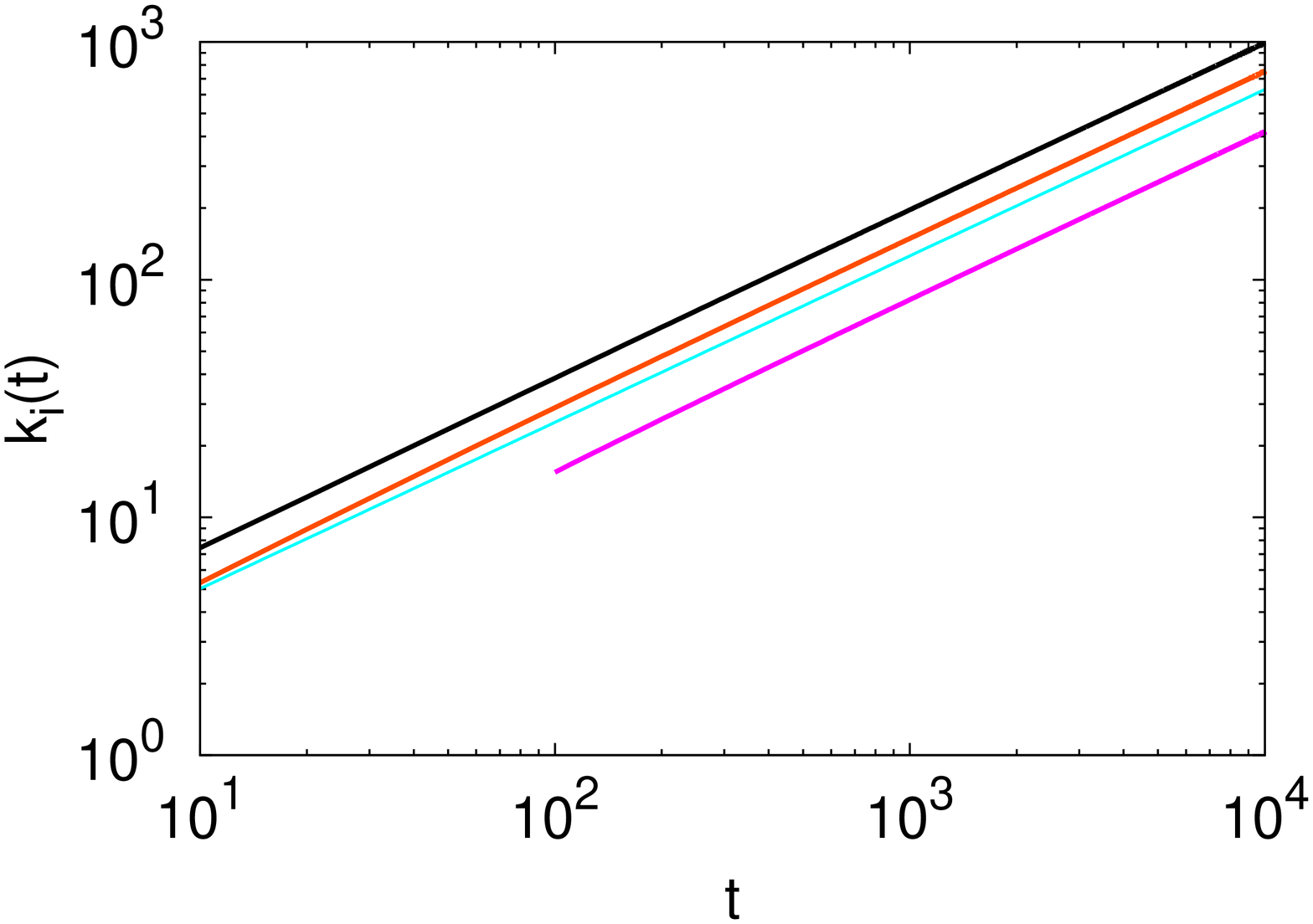}
     \begin{center} (a) $\delta=0.3$ \end{center}
 \end{minipage} 
 \hfill 
 \begin{minipage}{0.47\textwidth} 
   \centering
   \includegraphics[height=71mm,angle=-90]{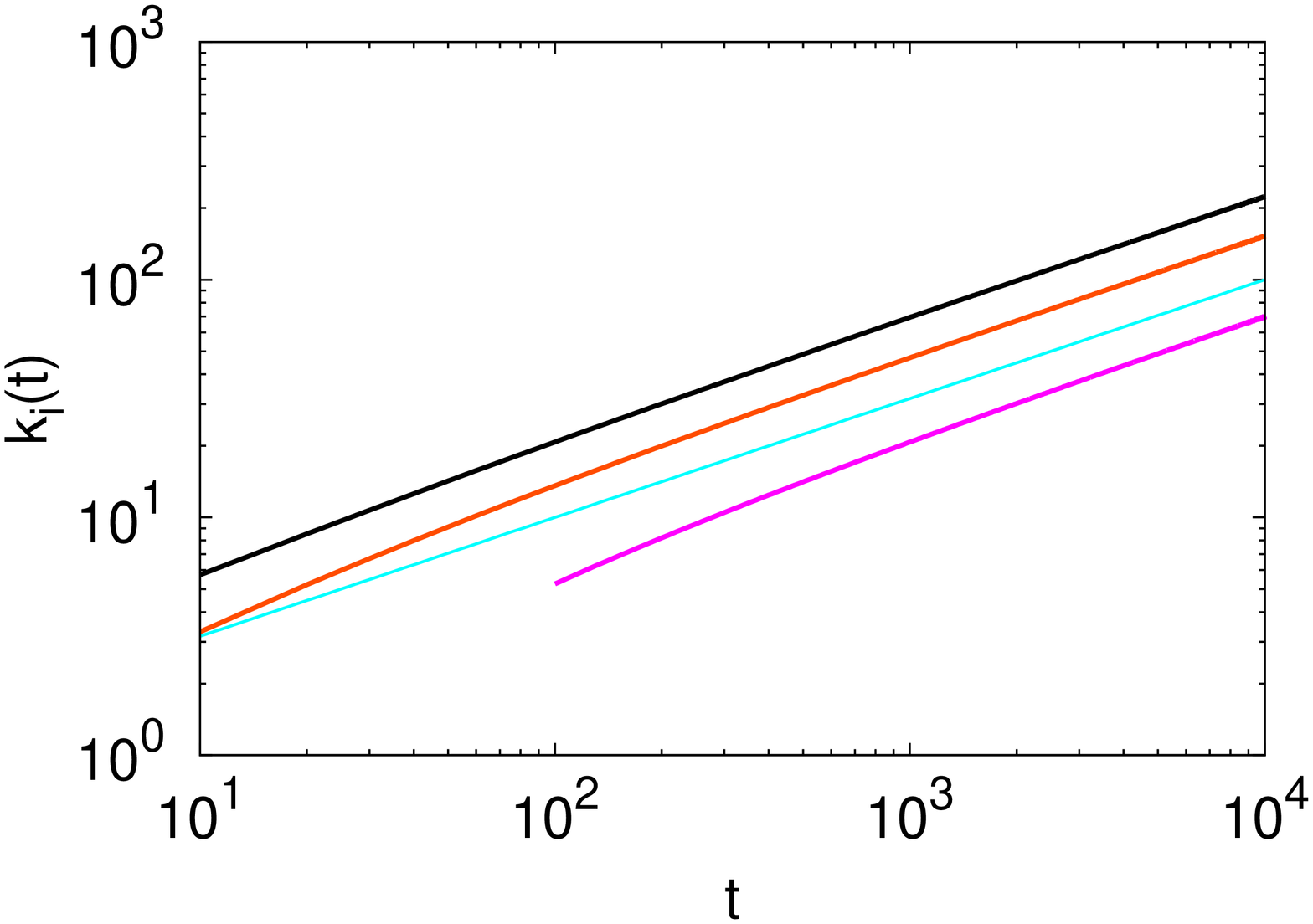}
     \begin{center} (b) $\delta=0.5$ \end{center}
 \end{minipage} 
 \hfill 
 \begin{minipage}{0.47\textwidth} 
   \centering
   \includegraphics[height=71mm,angle=-90]{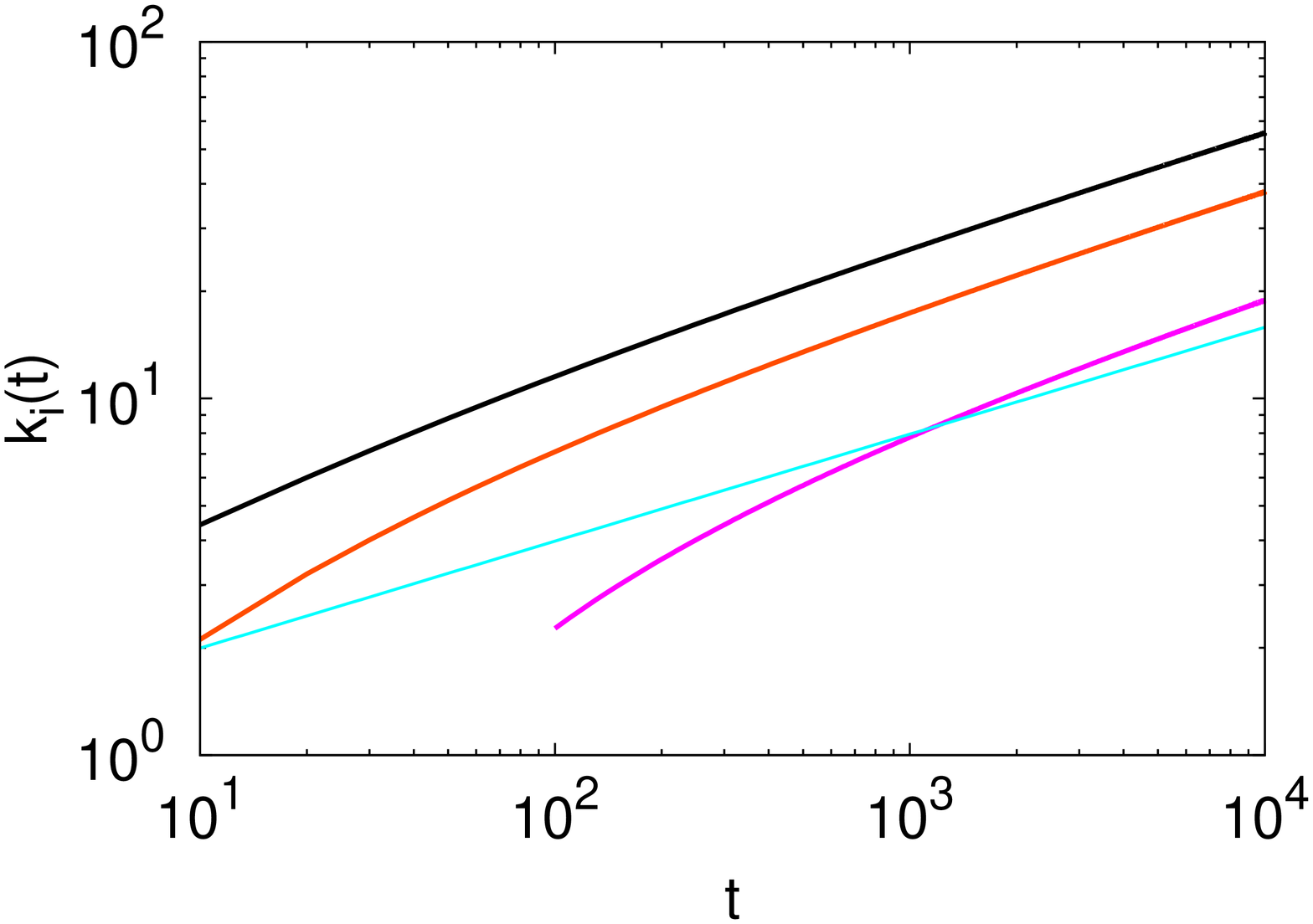}
     \begin{center} (c) $\delta=0.7$ \end{center}
 \end{minipage} 
 \hfill 
 \begin{minipage}{0.47\textwidth} 
   \centering
   \includegraphics[height=67mm,angle=-90]{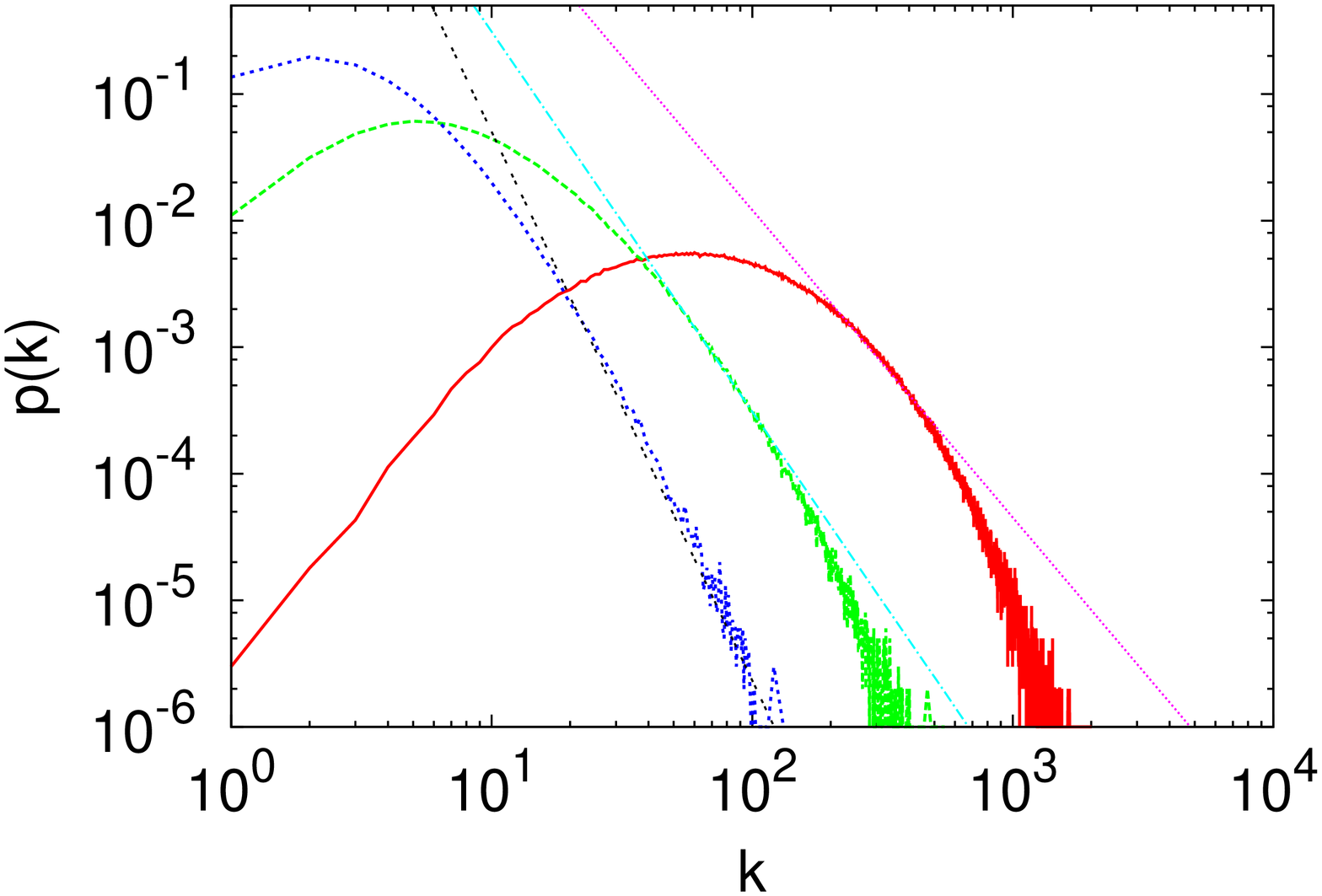}
     \begin{center} (d) $p(k)$  \end{center}
 \end{minipage} 
\caption{Results for the copying model.
(a)-(c) Time-courses $k_{i}(t)$ of degree for node $i=1, 10,$ and $100$ 
shown by black, orange, and magenta lines, respectively, 
at the typical values of rate $\delta$ of link deletion.
The cyan line guides the estimated slope $1 - \delta$ of 
$k_{i}(t) \sim t^{1 - \delta}$. 
(d) Degree distribution $p(k)$ in the cases of 
$\delta = 0.3, 0.5$, and $0.7$ shown by red, green, and blue lines, 
respectively.
The dotted magenta, cyan, and gray lines show the estimated power-law
distributions $k^{-\left(1+\frac{1}{1-\delta}\right)}$.}
\label{fig_copying}
\end{figure}

Figure \ref{fig_copying}(a)-(c) shows the time-course of 
$k_{i}(t) \sim t^{1-\delta}$ in the cases of 
$\delta = 0.3$, $0.5$, and $0.7$, respectively, 
averaged over $100$ samples. 
The black, orange, and magenta lines are the numerical 
results of Eqs. (\ref{eq_DE_mut_ki})(\ref{eq_DE_mut_kn}) 
for the node $i = 1$, $10$, and $100$.
The cyan line guides the estimated slope of 
$1-\delta$ in log-log plot.
It fits to the lines of $k_{i}(t)$ for a large $t$.
Moreover, as shown in Fig. \ref{fig_copying}(d), 
Eq. (\ref{eq_approx_pk}) gives a good approximation at the size 
$n =10^{4}$.
The red, green, and blue lines show the degree distributions for 
$\delta = 0.3$, $0.5$, and $0.7$, respectively.
The magenta, cyan, and gray dashed lines guides the corresponding 
slopes of $1 + \frac{1}{1-\delta}$ for these $\delta$ in the fitting
to the tails.

\subsection{Copying model with positive degree-degree correlations}
In this subsection, 
we emphasize that our approach is effective 
through the numerical estimation, 
even when an analytic derivation is intractable. 

We consider a copying model with positive degree-degree correlations 
based on a cooperative generation mechanism by linking 
homophily, 
in which densely connected cores among high degree nodes emerge 
\cite{Hayashi14}.
In more detail, 
the difference to the previously mentioned copying model is that 
the $n$-th new node links to the neighbor nodes $j$ of 
a randomly chosen node $i$ with a probability 
$(1-\delta) f((1-\delta)k_{i}(t-1), k_{j}(t-1))$
from existing nodes in the network.
Such a function 
\[
  f(x, y) \stackrel{\rm def}{=} \frac{1}{1 + a | x - y |},
\]
is necessary to enhance the 
degree-degree correlations, and $a > 0$ is a parameter \cite{Wu11}. 
Since the degree of new node is unknown in advance 
due to the stochastic process, 
it is temporary set as $(1-\delta) k_{i}(t-1)$. 

Thus, instead of Eq. (\ref{eq_a_ni_copying}), 
we substitute 
\begin{equation}
  a_{ni} \stackrel{\rm def}{=} 
  \frac{1 + (1-\delta)\sum_{j=1}^{n-1} a_{ji} 
    \times f((1-\delta)k_{i}(t-1), k_{j}(t-1))}{n-1},
\label{eq_a_ni_correl}
\end{equation}
for Eqs. (\ref{eq_DE_general_ki})(\ref{eq_DE_general_kn}). 

\begin{figure}[htp]
 \begin{minipage}{0.47\textwidth} 
   \centering
   \includegraphics[height=67mm,angle=-90]{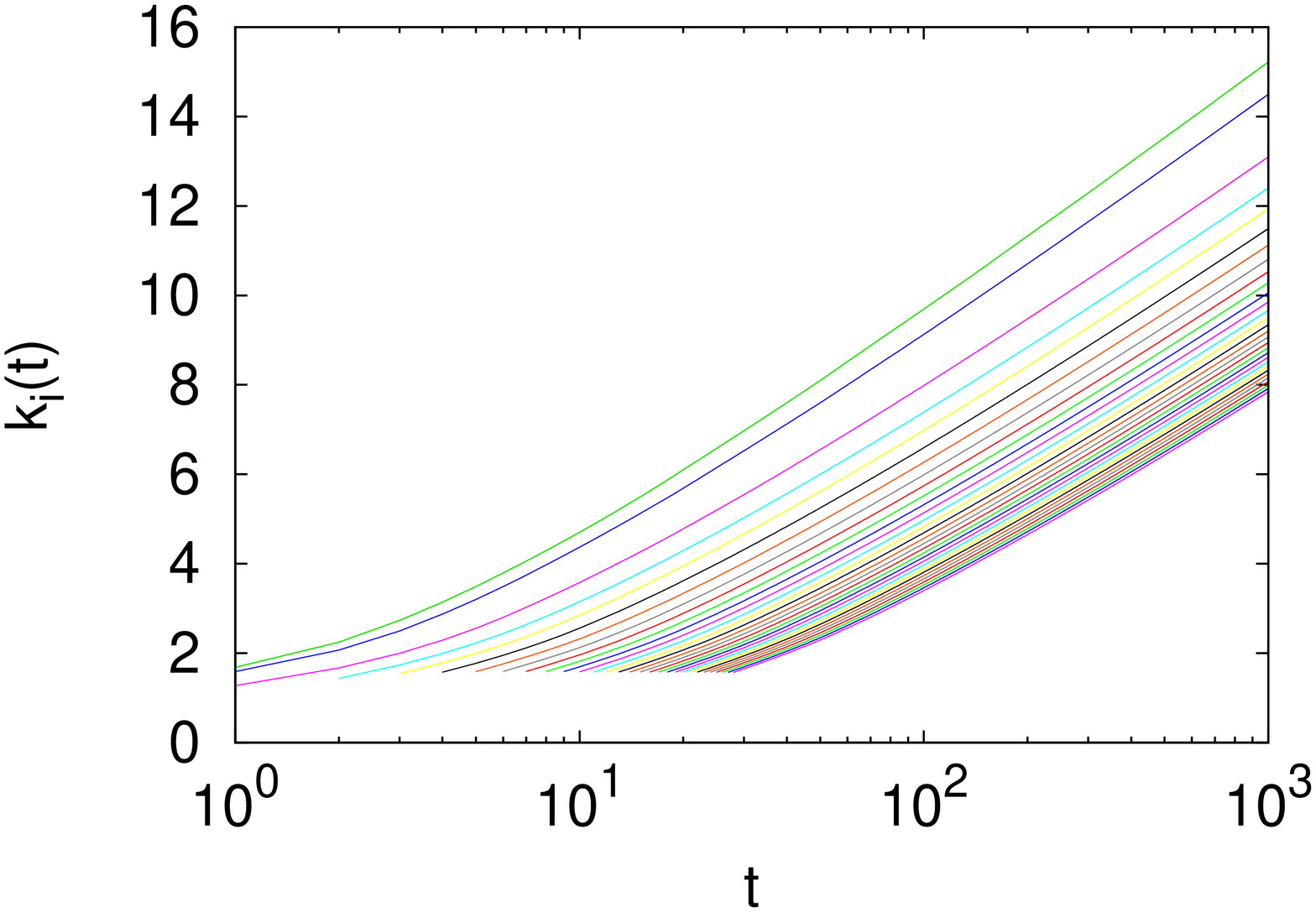}
     \begin{center} (a) $\delta=0.3$ \end{center}
 \end{minipage} 
 \hfill 
 \begin{minipage}{0.47\textwidth} 
   \centering
   \includegraphics[height=67mm,angle=-90]{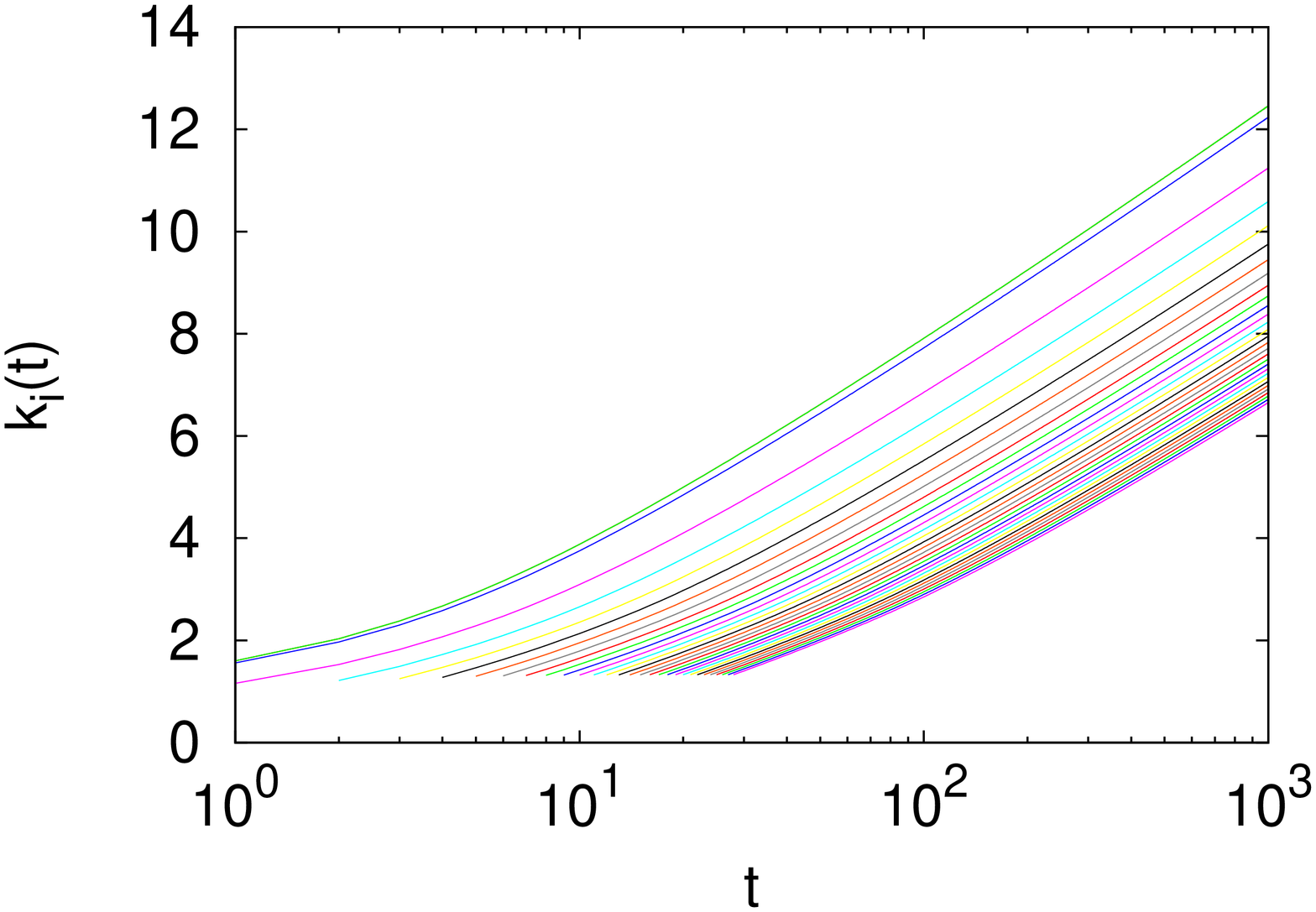}
     \begin{center} (b) $\delta=0.5$ \end{center}
 \end{minipage} 
 \hfill  
 \begin{minipage}{0.47\textwidth} 
   \centering
   \includegraphics[height=67mm,angle=-90]{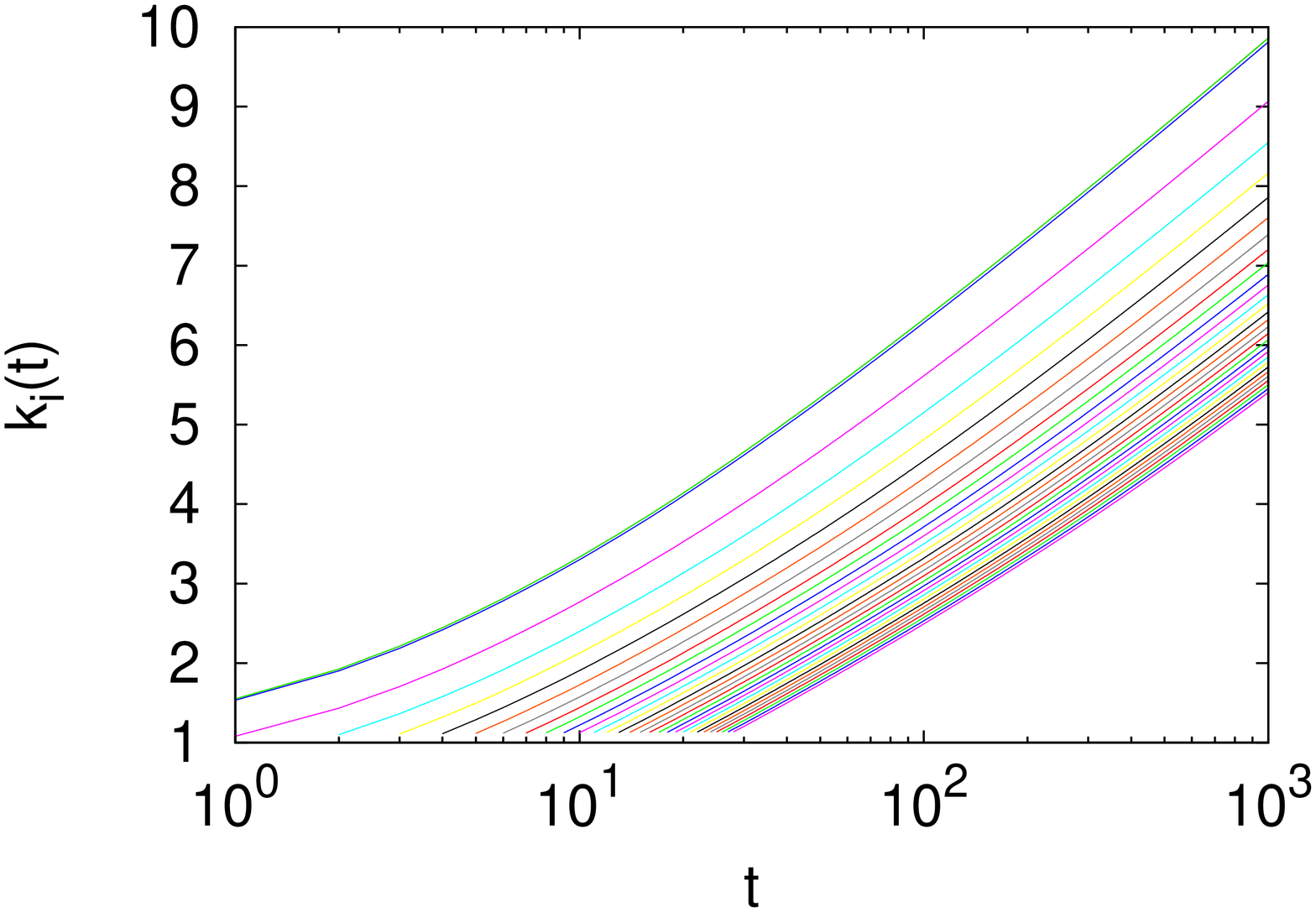}
     \begin{center} (c) $\delta=0.7$ \end{center}
 \end{minipage} 
\caption{Parallel curves of time-course $k_{i}(t)$ of degree 
in the copying model with 
positive degree-degree correlations and a rate $\delta$ of link deletion. 
The color lines from top to bottom correspond to the nodes 
$i=1$ or $2, 3, 4, \ldots, 30$. Note that each node $i$ is born at time 
$t_{i}=i-2$.}
\label{fig_para_correl}
\end{figure}

Although the theoretical analysis 
of Eqs. (\ref{eq_DE_general_ki})(\ref{eq_DE_general_kn}) 
for Eq. (\ref{eq_a_ni_correl}) is difficult, 
the iterative calculations are possible numerically.
When we assume $\beta k_{i}(t) \sim \log(t)$, 
we derive an exponential distribution as follows.
From $t \sim e^{\beta k_{i}(t)}$, we obtain 
\[
  \frac{t_{i}}{t} = \frac{e^{\beta k_{i}}}{e^{\beta k_{i}(t)}},
\]
\[
  p(k_{i}(t) < k) = 
  p(t_{i} > \frac{e^{\beta k_{i}}}{e^{\beta k}} t )
    = \left( 1 - \frac{e^{\beta k_{i}}}{e^{\beta k}} \right) 
    \frac{t}{N_{0}+t},
\]
\[
  p(k) = \frac{\partial p(k_{i}(t) < k)}{\partial k}
  \sim e^{- \beta k},
\]
under the invariant ordering (\ref{eq_ordering}) in the parallel 
curves shown as Fig. \ref{fig_para_correl}.

\begin{figure}[htp]
 \begin{minipage}{0.47\textwidth} 
   \centering
   \includegraphics[height=67mm,angle=-90]{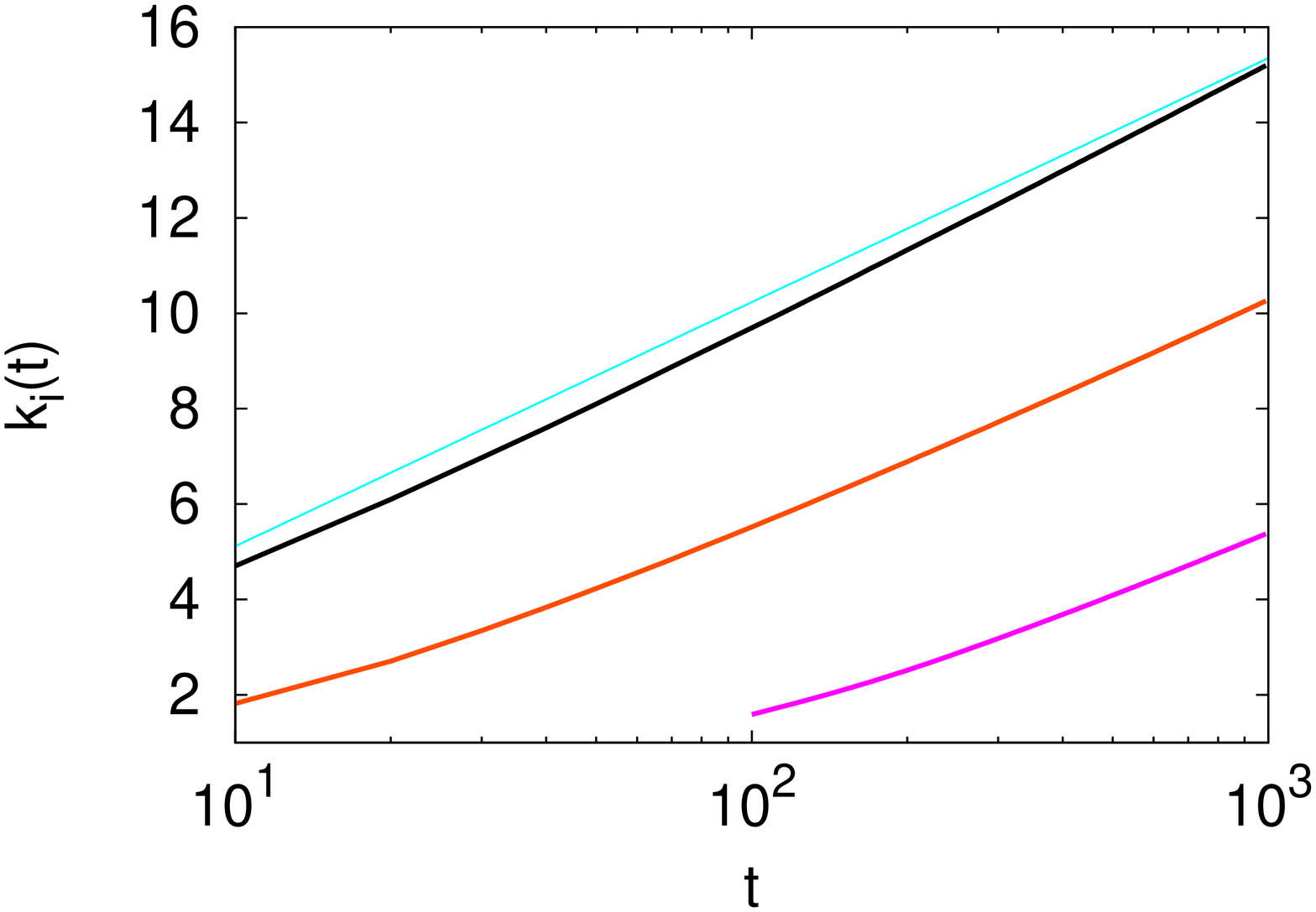}
     \begin{center} (a) $\delta=0.3$ \end{center}
 \end{minipage} 
 \hfill 
 \begin{minipage}{0.47\textwidth} 
   \centering
   \includegraphics[height=67mm,angle=-90]{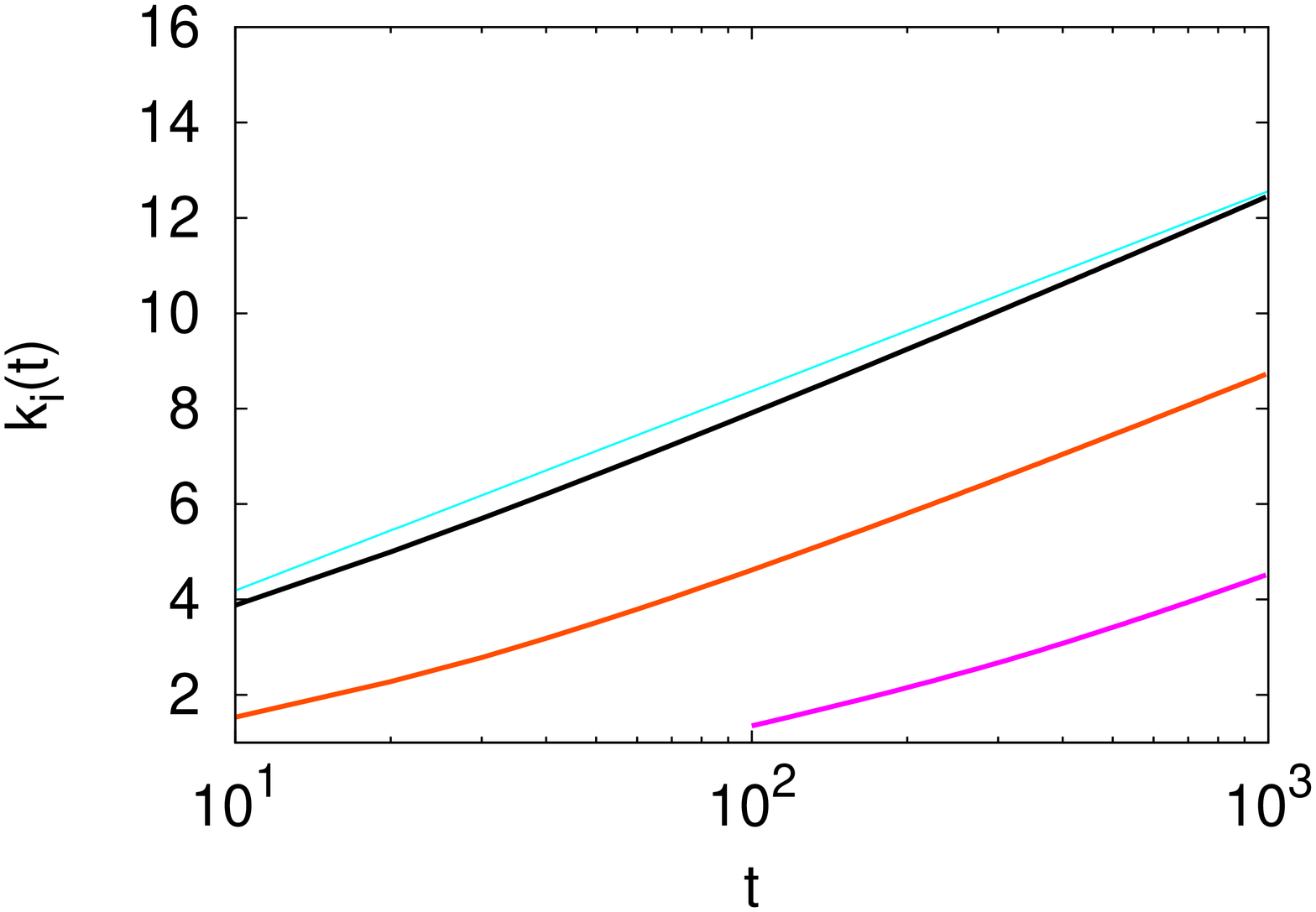}
     \begin{center} (b) $\delta=0.5$ \end{center}
 \end{minipage} 
 \hfill 
 \begin{minipage}{0.47\textwidth} 
   \centering
   \includegraphics[height=67mm,angle=-90]{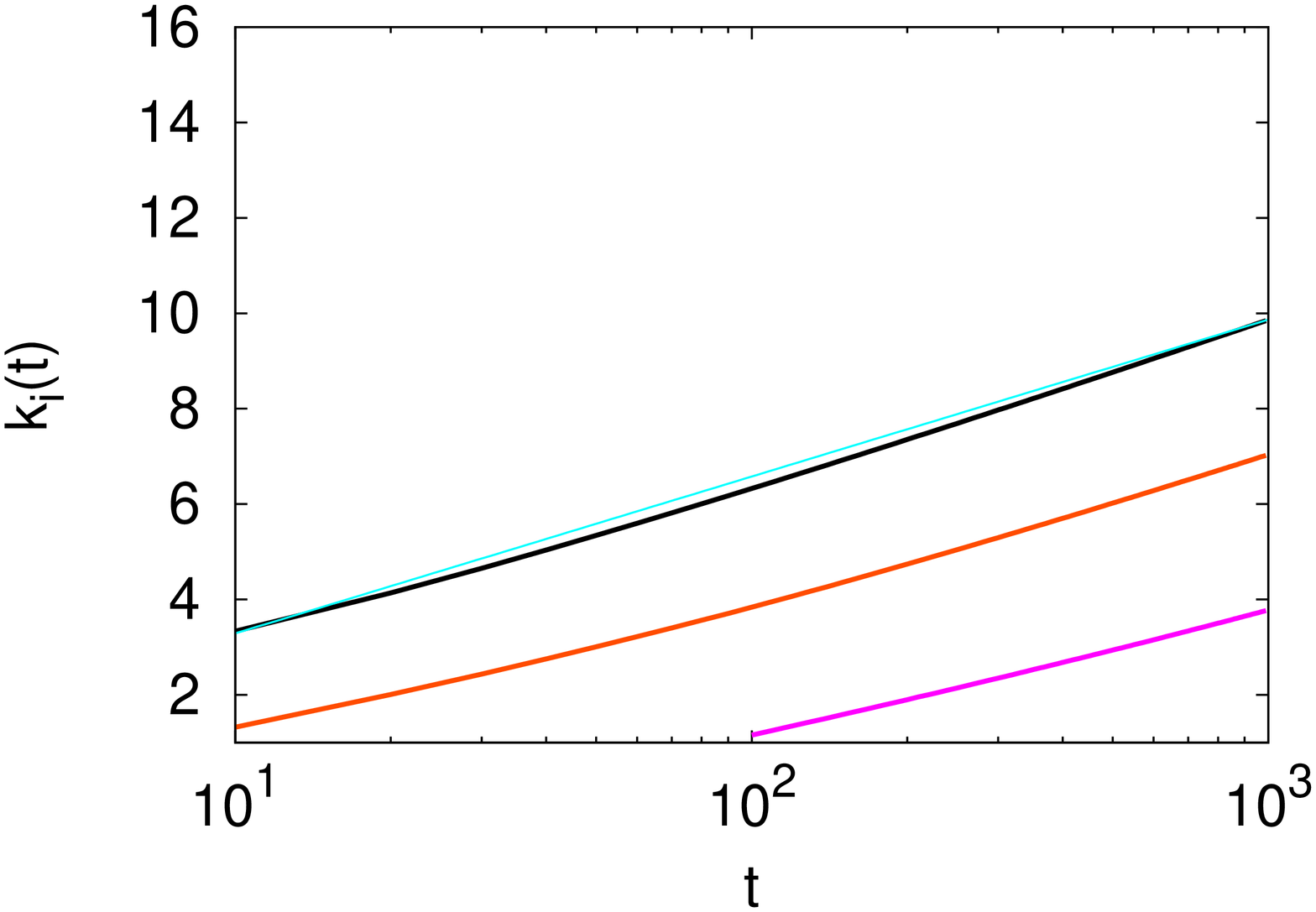}
     \begin{center} (c) $\delta=0.7$ \end{center}
 \end{minipage} 
 \hfill 
 \begin{minipage}{0.47\textwidth} 
   \centering
   \includegraphics[height=67mm,angle=-90]{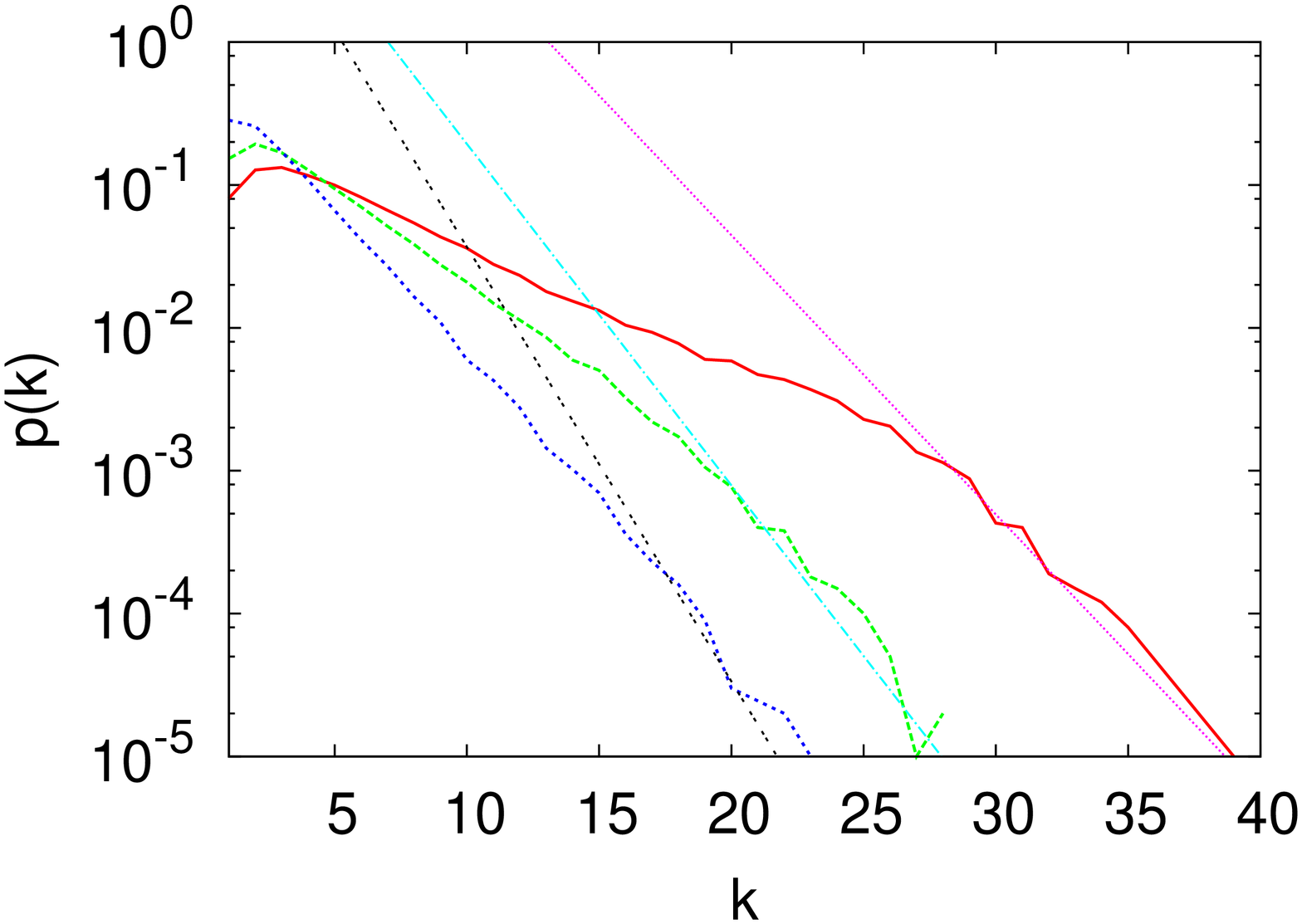}
     \begin{center} (d) $p(k)$  \end{center}
 \end{minipage} 
\caption{Results for the copying model with positive degree-degree correlations.
(a)-(c) Time-courses $k_{i}(t)$ of degree for node $i=1, 10,$ and $100$ 
shown by black, orange, and magenta lines, respectively, 
at the typical values of rate $\delta$ of link deletion.
The cyan line guides the estimated slope $1 / \beta$ of 
$k_{i}(t) \sim \log (t) / \beta$ in semi-log plot.
(d) Degree distribution $p(k)$ in the cases of 
$\delta = 0.3, 0.5$, and $0.7$ shown by red, green, and blue lines, 
respectively.
The dotted magenta, cyan, and gray lines show the estimated 
exponential distributions $e^{- \beta k}$.}
\label{fig_correl}
\end{figure}

Figure \ref{fig_correl}(a)-(c) shows that 
$k_{i}(t)$ denoted by black, orange, and magenta lines for 
$i = 1$, $10$, and $100$ is approximated by $\log(t)$ in the copying
model with degree-degree correlations.
The cyan lines guide the estimated slopes 
$\beta = 2.222$, $1.818$, and $1.428$ 
for $\delta = 0.3$, $0.5$, and $0.7$, respectively, 
in the numerical fittings for the iterative calculations of 
Eqs. (\ref{eq_DE_general_ki})(\ref{eq_DE_general_kn})(\ref{eq_a_ni_correl}). 
Thus, as shown in Fig. \ref{fig_correl}(d), 
the tails in $p(k)$ denoted by red, green, and blue lines 
for $\delta = 0.3$, $0.5$, and $0.7$ are approximated by 
$e^{-\beta k}$ shown as magenta, cyan, and gray dashed lines
at the size $n =10^{3}$. 
Note that 
$p(k)$ is only slightly deviated but the exponential part 
is remained by adding shortcut links in order to self-organize 
a robust onion-like structure 
in the incrementally growing network \cite{Hayashi14}.

\subsection{More general case}
We consider the asymptotic behavior of the node degrees 
for a general case in growing networks. 
When the time-course $k_{i}(t)$ of degree follows 
a monotone increasing function $g(t)$ of time $t$, 
there exists the inverse function 
$h(k) \stackrel{\rm def}{=} g^{-1}(k) = t$.
It is possible that 
the time-course $k_{i}(t)$ is an average of observed real data 
for node $i$, e.g. born at time $t-2$. 
From 
\[
  \frac{t_{i}}{t} = \frac{h(k_{i})}{h(k)}, 
\]
we have 
\[
   p(k_{i}(t) < k) = 
   p\left(t_{i} > \frac{h(k_{i})}{h(k)} t\right).
\]
Then, on the assumption of the invariant ordering (\ref{eq_ordering}) 
in parallel curves of $\{ k_{i}(t) \}$, 
we derive 
\[
  p(k) = \frac{\partial p(k_{i}(t) < k)}{\partial k} 
  = \frac{\partial}{\partial k} \left( 1 - 
  \frac{{\rm const.}}{h(k)} \right) \frac{t}{N_{0} + t} 
  \sim \frac{h'(k)}{h^{2}(k)}.
\]
where $h'(k)$ denote the derivative of $h(k)$ by the variable 
$k$.
We should remark that various 
degree distributions of non-power-law may appear depending on 
the shape of monotone increasing function $k_{i}(t)$
according to what type of generation in growing networks.

\section{Analytic deviation for the invariant 
ordering of node degrees}
We derive the invariant ordering (\ref{eq_ordering}) 
at any time $t$ within parallel curves of monotone increasing functions
for the D-D and copying models discussed in subsections 
3.2 and 3.3.
In the following, 
we use double mathematical induction for node index $i$ and time $t$.

Once $k_{i}(t-1) - k_{j}(t-1) > 0$
is satisfied 
for $1$ or $2 \leq i < j \leq n-1$ at $t = n-2 \geq 1$, 
\begin{eqnarray}
  k_{i}(t) - k_{j}(t) & = &
  (k_{i}(t-1) - k_{j}(t-1)) + (a_{in} - a_{jn}) \nonumber \\
  & = & \left( 1 + \frac{1 - \delta}{n-1} \right) 
  (k_{i}(t-1) - k_{j}(t-1)) > 0, \label{eq_assump}
\end{eqnarray}
is obtained from Eqs. 
(\ref{eq_DE_general_ki})(\ref{eq_DE_D-D_ki})(\ref{eq_DE_mut_ki}).
Here, 
from Eqs. (\ref{eq_DE_D-D_ki})(\ref{eq_DE_D-D_kn}) at $n=3$ 
with the initial condition $k_{1}(0) = k_{2}(0) = 1$, 
we obtain 
\[
  k_{1}(1) = k_{2}(1) = 1 + \frac{1 - \delta}{2}, 
  \;\;\; k_{3}(1) = 1 - \delta < k_{1}(1) = k_{2}(1).
\]
From Eqs. (\ref{eq_DE_mut_ki})(\ref{eq_DE_mut_kn}) at $n=3$
with the same initial condition, we also obtain 
\[
  k_{1}(1) = k_{2}(1) = 2 - \frac{\delta}{2}, \;\;\; 
  k_{3}(1) = 1 + \frac{1 - \delta}{2}(1+1) 
  < k_{1}(1) = k_{2}(1).
\]
Since the assumption in Eq. (\ref{eq_assump}) is satisfied for 
$i=1$ or $2$ and $j=3$ at $t=2$, we have 
\begin{eqnarray}
  k_{3}(2) & < & k_{2}(2) = k_{1}(2), \nonumber \\
  & \vdots \nonumber & \\
  k_{3}(t) & < & k_{2}(t) = k_{1}(t), \label{eq_induct1}
\end{eqnarray}
by applying Eq. (\ref{eq_assump}) recursively for $t \geq 2$.

On the other hand, 
from Eqs. (\ref{eq_DE_D-D_ki})(\ref{eq_DE_D-D_kn}), we rewrite 
\[
  k_{n}(t) = \frac{1 - \delta}{n-1} \left( 
  \sum_{i=1}^{n-2} k_{i}(t-1) + k_{n-1}(t-1) \right), 
\]
\[
  k_{n-1}(t) = \left(1 + \frac{1 - \delta}{n-1} \right) 
  k_{n-1}(t-1), 
\]
\[
  k_{n-1}(t-1) = \frac{1 - \delta}{n-2} 
  \sum_{i=1}^{n-2} k_{i}(t-2),
\]
at $t=n-2$, and 
\[
  k_{i}(t-1) = \left(1 + 
  \frac{1 - \delta}{n-2} \right) k_{i}(t-2),
\]
for $i = 1, 2, \ldots, n-2$. 
Then, we have 
\begin{equation}
  k_{n-1}(t) - k_{n}(t) = \frac{1 - \delta}{n-2} 
  \left( 1 - \frac{n - 1 - \delta}{n-1} \right)
  \sum_{i=1}^{n-2} k_{i}(t-2) > 0. \label{eq_delta_kn1}
\end{equation}

Similarly, 
from Eqs. (\ref{eq_DE_mut_ki})(\ref{eq_DE_mut_kn}),  we rewrite 
\[
  k_{n}(t) = 1 + \frac{1 - \delta}{n-1} \left( 
  \sum_{i=1}^{n-2} k_{i}(t-1) + k_{n-1}(t-1) \right), 
\]
\[
  k_{n-1}(t) = \frac{1}{n-1} 
  + \left( 1 + \frac{1 - \delta}{n-1} \right) k_{n-1}(t-1), 
\]
\[
  k_{n-1}(t-1) = 1 + \frac{1 - \delta}{n-2} 
  \sum_{i=1}^{n-2} k_{i}(t-2),
\]
at $t=n-2$, and
\[
  k_{i}(t-1) = \frac{1}{n-2} 
  + \left(1 + \frac{1 - \delta}{n-2} \right) k_{i}(t-2), 
\]
for $i = 1, 2, \ldots, n-2$. 
Then, we also have 
\begin{equation}
  k_{n-1}(t) - k_{n}(t) = 
  \frac{\delta}{n-1} + \frac{1 - \delta}{n - 2}
  \left( 1 - \frac{n - 1- \delta}{n-1} \right)
  \sum_{i=1}^{n-2} k_{i}(t-2) > 0. \label{eq_delta_kn2}
\end{equation}

By applying (\ref{eq_assump}) recursively for $t > n-2$ 
after substituting Eq. (\ref{eq_delta_kn1}) or (\ref{eq_delta_kn2})
to the right-hand side of Eq. (\ref{eq_assump}), 
we obtain 
\begin{eqnarray}
  k_{3}(t) - k_{4}(t) & > & 0, \nonumber \\
  k_{4}(t) - k_{5}(t) & > & 0, \nonumber \\
  & \vdots & \nonumber \\
  k_{n-1}(t) - k_{n}(t) & > & 0. \label{eq_induct2}
\end{eqnarray}
From Eqs. (\ref{eq_induct1})(\ref{eq_induct2}), 
we obtain the ordering (\ref{eq_ordering}) after all.

Next, we consider the existing condition of 
the ordering (\ref{eq_ordering}) 
within parallel curves of $\{ k_{i}(t) \}$
for a general case of growing networks  
discussed in subsection 3.5.

From Eq. (\ref{eq_DE_general_ki}) 
for $1$ or $2 \leq i < j \leq n-1$ at $t = n-2 \geq 2$, 
we have 
\begin{equation}
  k_{i}(t) - k_{j}(t) =
  (k_{i}(t-1) - k_{j}(t-1)) + (a_{in} - a_{jn}).
  \label{eq_general_ij}
\end{equation}
If older nodes tend to get more links, 
\begin{equation}
  a_{in} > a_{jn} \label{eq_cond_general} 
\end{equation}
hold in the ensemble average of adjacency matrix.
Then, we remark 
$k_{3}(t) < k_{2}(t) = k_{1}(t)$
from Eq. (\ref{eq_cond_general}) and 
$k_{3}(1) = k_{3}(0) + a_{13} + a_{23} 
< k_{1}(1) = k_{1}(0) + a_{31}$ 
or $k_{2}(1) =  k_{2}(0) + a_{32}$ 
with the initial values $k_{3}(0) = 0$ and 
$k_{1}(0) = k_{2}(0) =1$.

On the other hand, 
from Eqs. (\ref{eq_DE_general_ki})(\ref{eq_DE_general_kn}), 
we derive 
\[
  k_{n-1}(t) - k_{n}(t) 
  = \sum_{l=1}^{n-2} (a_{(n-1)l} - a_{nl}). 
\]
If Eq. (\ref{eq_cond_general}) hold 
with $a_{ij} = a_{ji}$, for $n \geq 4$ we have 
\begin{equation}
  k_{n-1}(t) - k_{n}(t) > 0. \label{eq_general_n}
\end{equation}
Therefore, on the condition (\ref{eq_cond_general}), 
we obtain the ordering (\ref{eq_ordering}) from 
Eqs. (\ref{eq_general_ij})(\ref{eq_general_n}).

\section{Conclusion}
We have proposed the explicit representation 
by the ensemble average of adjacency matrix 
over samples of growing networks. 
The important point is that 
the adjacency matrix is averaged in advance 
before calculating a characteristic 
quantity about the topological structure for each sample.
The ensemble average has been applied to some network models: 
BA \cite{Barabasi99b},  D-D \cite{Satorras03,Sole02}, and 
copying \cite{Hayashi14} models for investigating the 
degree distributions in the asymptotic behavior 
by using the theoretical and numerical analysises for
difference equations and the corresponding     
continuous-time approximation of differential equations
with variables $t$ and $k_{i}(t)$.
We have derived 
$k_{i}(t) \sim t^{1-\delta}$ and 
$p(k) \sim k^{-(1+\frac{1}{1-\delta})}$ 
for the D-D and copying models under the invariant ordering 
of degrees which is supported in randomly grown networks 
\cite{Callaway01}. 
Moreover, 
for the copying model with positive degree-degree correlations, 
we have shown that 
the numerical calculations of the difference equation 
give a good approximation of an estimated 
exponential distribution, 
even when an analytic derivation is intractable. 
The copying model with positive degree-degree correlations 
is related to the self-organization \cite{Hayashi14} 
of robust onion-like networks 
\cite{Herrmann11,Schneider11,Tanizawa12}.

Our approach may be also 
applicable to data analysis for social networks, 
when the observed time-course of degree is a monotone increasing 
function like power-law or logarithm in the average over samples 
by ignoring short-time fluctuations, 
and a node index $i$ or $j$ represents an ordering of 
its birth time.
This expectation is supported as follows. 
It is helpful for grasping a trend 
to study the average behavior of many 
users (nodes) added at a same (sampling interval) time into 
a network community. 
Random growth in a social network probably corresponds to 
encountered chances among people. 
Moreover, as time goes by, 
the number of his/her friends for a member of social networks 
is usually increasing. 
It is natural that 
the connections to friends are maintained. 
However, we must consider the effect of rewirings between 
old members in the definition of adjacency matrix.
Also from a practical viewpoint, 
$O(n^{2})$ space to store an adjacency matrix 
may cause a problem for dig data.

From the conventional analysis for special network models 
to a general framework, 
the representation by the ensemble average will open 
a door for investigating the characteristic quantities, 
e.g. node degrees in growing networks. 
In particular, 
the time-course of a quantity depending on the birth time 
of node is considered as a key point. 
The discussion about other quantities such as 
clustering coefficient or the number of paths of a given length 
requires further studies of how to analyze 
the average behavior related to the transitivity.

\section*{Acknowledgments}
The author would like to thank anonymous reviewers 
for their valuable comments.
This research is supported in part by
a Grant-in-Aid for Scientific Research in Japan, No. 25330100.


\begin{thebibliography}{62}
 \bibitem{Albert00}
   Albert,~R., Jeong,~H. and Babar\'{a}si,~A.-L. 
   Error and attack tolerance of complex networks. 
   {\em Nature} 406: pp.36--44, (2000).
 \bibitem{Barabasi99a}
   Babar\'{a}si,~A.-L. and Albert,~R. 
   Emergence of scaling in random networks.
   {\em Science} 286: pp.509--512, (1999).
 \bibitem{Barabasi99b}
   Babar\'{a}si,~A.-L., Albert,~R. and Jeong,~H. 
   Mean-field theory for scale-free random networks. 
   {\em Physica A} 272: pp.173--187, (1999).
 \bibitem{Callaway01}
   Callaway,~D.S., Hopcroft,~J.E., Kleinberg,~J.M., Newman,~M.E.J. and Strogatz,~S.H. 
   Are randomly grown graphs really random ?. 
   {\em Physical Review E} 64: pp.041902, (2001).
 \bibitem{Hayashi14}
   Hayashi,~Y. 
   Growing Self-organized Design of Efficient and Robust Complex Networks.
   {\em IEEE Xplore Digital Library} 
   http://dx.doi.org/10.1109/SASO2014.17, 
   Proc. of 2014 IEEE 8th Int.
   Conf. on SASO: Self-Adaptive and Self-Organizing Systems 2014, 
   pp.50--59. arXiv:physics/1411.7719, (2014). 
 \bibitem{Herrmann11}
   Herrmann,~H.J., Schneider,~C.M., Moreira,~A.A., Andrade Jr.~J.S., and 
   Havlin,~S. 
   Onion-like network topology enhances robustness against malicious attacks.
   {\em Journal of Statistical Mechanics} P01027, (2011).
 \bibitem{Kim02}
   Kim,~J., Krapivsky,~P.L. and Redner,~S. 
   Infinite-order percolation and giant fluctuations 
   in a protein interaction networks. 
   {\em Physical Review E} 66: pp.055101(R), (2002). 
 \bibitem{Newman03a}
   Newman,~M.E.J. 
   Assortative Mixing in Networks. 
   {\em Physical Review Letters} 89(20): pp.208701, (2003).
 \bibitem{Newman10}
   Newman,~M.E.J. 
   Networks -An Introduction. Oxford University Press, (2010). 
 \bibitem{Satorras03}
   Pastor-Satorras,~R., Smith,~E. and Sole,~R.V. 
   Evolving protein interaction networks through gene duplication. 
   {\em Journal of Theoretical Biology} 222(2): pp.199--210, (2003).
 \bibitem{Schneider11}
   Schneider,~C.M., Moreira,~A.A., Andrade Jr.~J.S., Havlin,~S. 
   and Herrmann,~H.J. 
   Mitigation of malicious attacks on networks.
   {\em Proceedings of the National Academy of Sciences of the United States of America} 
   810(10): pp.3838--3841, (2011).
 \bibitem{Sole02}
   Sole,~R.V., Pastor-Satorras,~R., Smith,~E. and Kepler,~T.B. 
   A model of large-scale proteome evolution.
   {\em Advances in Complex Systems} 5(1): pp.43--54, (2002).
 \bibitem{Tanizawa12}
   Tanizawa,~T., Havlin,~S. and Stanley,~H.E. 
   Robustness of onionlike correlated networks against targeted attacks.
  {\em Physical Review E} 85: pp.046109, (2012).
\bibitem{Wu11}
  Wu,~Z.-X. and Holme,~P. 
  Onion structure and network robustness.
  {\em Physical Review E} 81: pp.026116, (2011).
\end{thebibliography}
\end{document}